\documentclass[twocolumn]{aastex631}
\usepackage{soul}
\usepackage{gensymb}
\usepackage{array,multirow}
\usepackage{ragged2e}
\usepackage{xcolor}

\begin{document}

\title{An Explainable Deep-learning Model of Proton Auroras on Mars}

\author[0000-0001-7927-2727]{Dattaraj B. Dhuri}
\affiliation{Center for Astrophysics and Space Science \\
New York University Abu Dhabi \\
PO Box 129188, Saadiyat Island, Abu Dhabi, UAE}

\author{Dimitra Atri}
\affiliation{Center for Astrophysics and Space Science \\
New York University Abu Dhabi \\
PO Box 129188, Saadiyat Island, Abu Dhabi, UAE}

\author[0000-0002-7872-1954]{Ahmed AlHantoobi}
\affiliation{Center for Astrophysics and Space Science \\
New York University Abu Dhabi \\
PO Box 129188, Saadiyat Island, Abu Dhabi, UAE}
\affiliation{Propulsion Department \\
Advanced Concepts - EDGE Group, Abu Dhabi, UAE}

\correspondingauthor{Dattaraj B. Dhuri}
\email{dbd7602@nyu.edu}

\begin{abstract}
Proton auroras are widely observed on the dayside of Mars, identified as a significant intensity enhancement in the hydrogen Lyman‐${\rm \alpha}$ (Ly-${\rm \alpha}$, 121.6 nm) emission between $\sim$ 110 – 150~km altitudes. Solar wind protons penetrating as energetic neutral atoms into Mars’ thermosphere are thought to be primarily responsible for these auroras. Recent observations of spatially localized ``patchy" proton auroras suggest a possible direct deposition of protons into Mars’ atmosphere during unstable solar wind conditions. Improving our understanding of proton auroras is therefore important for characterizing the solar wind interaction with Mars’ atmosphere. Here, we develop a first purely data-driven model of proton auroras using Mars Atmosphere and Volatile EvolutioN (MAVEN) in-situ observations and limb scans of Ly-${\rm \alpha}$ emissions between 2014 -- 2022. We train an artificial neural network (ANN) that reproduces individual Ly-${\rm \alpha}$ intensities and relative Ly-${\rm \alpha}$ peak intensity enhancements with a Pearson correlation of $\sim$ 94\% and $\sim$ 60\% respectively for the test data, along with a faithful reconstruction of the shape of the observed Ly-${\rm \alpha}$ emission altitude profiles. By performing a SHapley Additive exPlanations (SHAP) analysis, we find that solar zenith angle, solar longitude, ${\rm CO_2}$ atmosphere variability, solar wind speed and temperature are the most important features for the modeled Ly-${\rm \alpha}$ peak intensity enhancements. Additionally, we find that the modeled peak intensity enhancements are high for early local time hours, particularly near polar latitudes, as well as weaker induced magnetic fields. Through SHAP analysis, we also identify the influence of biases in the training data and interdependecies between the measurements used for the modeling, and an improvement on those aspects can significantly improve the performance and applicability of the ANN model. 
\end{abstract}


\section{Introduction} \label{sec:intro}
Auroras on Mars are observed as enhancements in FUV and EUV emissions on both night as well as dayside of Mars \citep{Bertaux2005,Schneider2015,Deighan2018}. Three distinct type of such auroras have been observed. The discrete auroras were the first auroras observed on Mars \citep{Bertaux2005}. They are typically caused by electrons moving from the dayside to the nighside along the closed crustal magnetic-field lines and are highly localized. In contrast, the diffuse electron auroras on Mars are observed during Solar Energetic Particle (SEP) events and are caused by higher energy electrons penetrating along the open magnetic-field lines across the planet \citep{Schneider2015}. In addition to discrete and diffuse auroras observed over the years, proton auroras are a relatively newly discovered phenomenon \citep{Deighan2018,Ritter2018,Chaffin2022} observed mainly on the dayside of the Mars. Both electron and proton auroras occur extremely frequently \citep{Hughes2019, Lillis2022} and therefore studying them can provide new insights into the complex interactions between the solar wind and weak crustal field of the planet and its surrounding plasma environment (see \citet{atri2022auroras} for a recent review).

Proton auroras are reported to be one of the most widely observed auroras on Mars, identified in ${\rm \sim 14\%}$ of observations \citep{Hughes2019} from Imaging UltraViolet Spectrograph (IUVS) \citep{McClintock2015} onboard the Mars Atmosphere and Volatile EvolutioN (MAVEN) spacecraft \citep{Jakosky2015}. On Mars, these auroras are thought to be caused primarily by a population of solar wind protons penetrating the Martian magnetosphere as hydrogen energetic neutral atoms (ENAs) \citep{Deighan2018}. These hydrogen ENAs are formed in the outer hydrogen corona of Mars via electron stripping and charge exchange. Once in the thermosphere, the ENAs undergo repeated charge exchange and collisions with the neutrals, and can emit Lyman-${\rm \alpha}$ (Ly-${\rm \alpha}$, 121.6 nm) radiation that is seen as proton auroras.

Using Ly-${\rm \alpha}$ emission profiles from MAVEN/IUVS, \citet{Deighan2018} first reported a proton aurora observation on Mars, characterized by a Ly-${\rm \alpha}$ intensity enhancement in the altitude range 120 -- 150~km. They showed that these emission enhancements are correlated with the observed penetrating proton flux \citep{Halekas2015}, and suggested that these auroras are thus triggered by hydrogen ENAs. Subsequently, \citet{Ritter2018} presented UV data from the Spectroscopy for the Investigation of the Characteristics of the Atmosphere of Mars (SPICAM) \citep{Bertaux2006} onboard Mars Express (MEX) to confirm the proton aurora observations. Recently, \citet{Chaffin2022} reported proton aurora observations in Ly-${\rm \alpha}$ and Ly-${\rm \beta}$ (102.6 nm) from Emirates Mars Ultraviolet Spectrometer (EMUS) \citep{Holsclaw2021} onboard the Emirates Mars Mission (EMM) \citep{Amiri2022}. EMUS captures a global synoptic view of UV emissions from Mars and has revealed that the proton aurora regions can be highly localized and ``patchy". Since the solar wind conditions that are known to trigger proton auroras are uniform across the dayside, these EMUS observations of ``patchy" proton auroras suggest existence of other mechanisms responsible for proton auroras on Mars (see \cite{Chaffin2022}).

Known physical processes involved in triggering proton auroras, namely the formation of ENAs and penetrating solar wind protons, form an important characteristic of the solar wind interaction with the Martian magnetosphere and escape of Mars' atmosphere. A comprehensive understanding of proton aurora occurrence characteristics and consequences for Mars' atmosphere evolution requires a thorough analysis of the influence of solar wind properties and the subsequent response of Mars' magnetosphere. \citet{Hughes2019} conducted a first statistical study of proton auroras observed by MAVEN/IUVS, to understand how the proton aurora occurrence rates and enhancements vary with the solar longitude (${\rm L_s}$), solar zenith angle (SZA), local time (lt) etc. among other factors. They found that the proton auroras occur between altitudes 110 -- 150~km, occurring at the lower altitudes especially around ${\rm L_s \sim 180\degree}$. Their analysis showed that the primary factors affecting proton auroras occurrence rates are ${\rm L_s}$ and SZA, with the highest occurrence rates and emission enhancements observed around southern summer solstice (${\rm L_s \sim 270\degree}$) and low SZAs. Hydrogen corona column densities above the bow shock and solar wind flux increase during this season (${\rm L_s \sim 270\degree}$) along with higher atmospheric temperatures and inflation of the lower atmosphere because of the increased dust activity. \citet{Hughes2019} suggested that these factors could contribute to a higher population of ENAs and their increased interaction with the lower atmosphere and therefore increased proton aurora occurrences and enhancements. \citet{Hughes2021T} further studied the influence of IMF upstream of the Martian bow shock, penetrating proton flux, dust and extreme solar activity on the Ly-${\rm \alpha}$ emission enhancements at the proton aurora peak altitudes. They reported a possible preference of proton aurora occurrence during radial IMF orientations. They also presented simple linear regression models of the orbit-averaged Ly-${\rm \alpha}$ emission enhancements, separately accounting for the cases of high dust activity and extreme solar activity, using MAVEN in-situ measurements of the orbit-averaged penetrating proton flux.

In this work, we attempt to explicitly model the influence of solar wind proton characteristics on Ly-$\alpha$ intensity enhancements. We consider the proton measurements of energy, density, temperature, velocity, and also in-situ magnetic fields obtained by MAVEN during its passage through the different regions of the magnetosphere in each orbit. We also consider the dependence on ${\rm CO_2}$ atmosphere density and Mars' crustal magnetic fields. These MAVEN measurements, referred to as features in this manuscript, are numerous and may have interdepedencies, plus a highly non-linear relationship with Ly-$\alpha$ emissions. We, therefore, develop an artificial neural network (ANN) model using these features as inputs to reproduce the observed Ly-$\alpha$ altitude profiles. Deep neural networks are extremely efficient in leveraging correlations from complex, high-dimensional large datasets to perform challenging tasks such as classification, regression, segmentation etc. \citep{Goodfellow2016}. The development, i.e. training or learning, of an ANN is posed as an optimization problem to obtain the ANN parameters that minimize a loss function or misfit between the modeled output and ground truth. Here, we demonstrate that the ANN is trained to learn dependencies between the input MAVEN observations and Ly-$\alpha$ emissions to accurately model the altitude profile observations.  

Modeling of the observed proton aurora Ly-${\rm \alpha}$ enhancements has been previously considered briefly in the original discovery paper by \citet{Deighan2018} and extensively in a recent ``multi-model campaign" by \citet{Hughes2023}. These physics-based modeling involved Monte Carlo simulations of proton/hydrogen precipitations in the Martian atmosphere, using the observed flux as input, and subsequent recreations of background-subtracted Ly-${\rm \alpha}$ altitude profiles using a radiative transfer model. \citet{Hughes2023} recreated and compared such simulations for MAVEN orbit \#4235 using four different Monte Carlo models and found solar wind particle flux and velocity as the primary variables affecting proton aurora. In contrast to these physics-based models, the ANN model considered in this work is purely data-driven and is used to recreate the Ly-${\rm \alpha}$ altitude profiles from many ($\sim$ 2000) MAVEN orbits simultaneously. However, unlike a physics-based model, a definite understanding of how a trained ANN uses given inputs for modeling is difficult and the reliability of an ANN model is therefore also questionable. Here we carry out a Shapley value analysis \citep{NIPS2017_7062} of the trained ANN to explore correlations of the input features used for modeling the Ly-$\alpha$ intensities. Through the Shapley analysis, we are able to identify possible biases and caveats in the data and modeling, recover previously known dependence of proton auroras on ${\rm L_s}$ and SZA, and also some new patterns in the data.

This paper is organized as follows. In Section~\ref{sec:Data}, we describe the selection and pre-processing of MAVEN data used for analysis. In Section~\ref{sec:Methods} we explain the ANN architecture and training methodology. In Section~\ref{sec:Results} we report our findings. Section~\ref{sec:performance} outlines the accuracy of our model. Section~\ref{sec:shap} presents a detailed analysis of Shapley values for different input features. In Section~\ref{sec:discussion} we summarize our results, discuss their implications, the scope of our model and possible improvements.

\section{Data} \label{sec:Data}
\begin{figure*}[ht!]
\includegraphics[width=\textwidth]{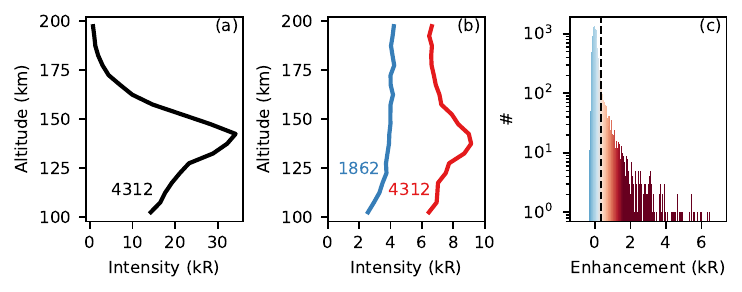}
\caption{Examples of MAVEN/IUVS limb scans showing the thermospheric altitude profiles of co2uvd emission (a) and Ly-${\rm \alpha}$ emissions (b). The Ly-${\rm \alpha}$ emissions show both a proton aurora case (red) and a non-proton aurora case i.e. the background dayglow emission (blue). The proton aurora profile shows the characteristic enhancement around 110 -- 150~km altitudes. Note that the co2uvd emission profiles between altitudes 130 -- 190~km are used in this study. The legend shows the orbit number for each profile. Distribution of the Ly-${\rm \alpha}$ enhancements within the data considered is shown in (c) (after \citep{Hughes2019}, with darker red also indicating increasing intensity of proton aurora enhancements. The dashed line marks an intensity enhancement threshold, used only as a reference, for defining the proton auroras as per \citet{Hughes2019}}.
\label{fig:iuvsExamples}
\end{figure*}
We use remote sensing and in-situ data from MAVEN between October 2014 --- April 2022 for developing an ANN model of the observed Ly-$\alpha$ emission altitude profiles. The data considered here covers almost four Martian years. 
Details of the analyzed data are as follows. 

\subsection{IUVS Limb-Scan Observations}
MAVEN/IUVS is a remote-sensing UV spectrograph monitoring the state of the Mars' upper atmosphere (110 -- 225~km). The IUVS wavelength range covers the FUV (110 -- 190~nm) and MUV (180 -- 340 nm) ranges. IUVS is thus sensitive to, among other emissions, Ly-$\alpha$ (121.6~nm) which is the focus of this study on proton auroras as well as ${\rm CO_2}$+UV Doublet band (288.3~nm and 289.6~nm, co2uvd) which we use as a proxy for the ${\rm CO_2}$ atmosphere density \citep{Deighan2018}. We use the publicly available Level 1C processed data products. During periapsis passes, IUVS operates in the limb-scan mode, when it records the altitude emission profiles between 100 -- 220~km \citep{McClintock2015}. Typically, 12 such scans are recorded for each periapsis passing segment lasting $\rm{\sim~23~min}$.

We only use limb-scan profiles that provide emissions for the full altitude range between 100 -- 200~km for Ly-$\rm{\alpha}$ and 130 -- 190~km for co2uvd. These altitude ranges are chosen to minimize the exclusion of limb-scan observations because of any missing data at lower or higher altitudes and still cover the region of Mars' thermosphere relevant for this study. The proton auroras in Ly-$\rm{\alpha}$ altitude scans are identified as an enhancement of emission intensity between 110 -- 150~km. Following \citet{Hughes2019}, we quantify the emission enhancement using an Enhancement Measure (EM) defined as the difference between the second highest intensity in the peak altitude range and the median value between 160 -- 190~km range. Figure~\ref{fig:iuvsExamples}a shows a sample co2uvd emission profile and Figure~\ref{fig:iuvsExamples}b shows samples of Ly-${\rm \alpha}$ profiles for a non-proton aurora (blue) observation and a proton aurora (red) observation. The latter shows a marked enhancement of Ly-${\rm \alpha}$ emission in the peak altitude range compared to the characteristic flat dayglow profile in the former. \citet{Hughes2019} defined an observation to be a proton aurora if EM exceeds ${\rm 0.5\sigma}$ than the mean EM, ${\rm \sigma}$ being the standard deviation of EMs in the data. Figure~\ref{fig:iuvsExamples}c shows the distribution of the enhancements across the data considered for the analysis, with the threshold highlighted by the dashed black line. \citet{Hughes2019} showed that the occurrence rate of proton auroras primarily depends on ${\rm L_s}$ and SZA. We, therefore, include these in our analysis along with other measurements latitude (lat), longitude (lon), and local time (lt) for each limb scan. The IUVS limb-scan data, with all altitude observations between the ranges specified above for Ly-${\rm \alpha}$ and co2uvd profiles, is available for 4130 orbits during the observation period considered for this study, out of total 16079 MAVEN orbits in this period.

\subsection{MAVEN in-situ measurements of protons and magnetic field} 
\begin{figure*}[ht!]
\includegraphics[width=\textwidth]{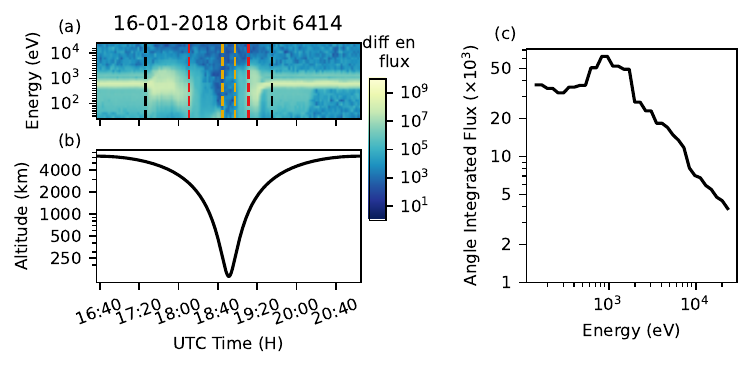}
\caption{Example of MAVEN/SWIA measurements for a sample orbit showing the variation in the proton energy spectra (a) and the MAVEN altitude (b). The bow-shock (black) and magnetic pile-up boundaries (red) marked by the dashed lines are from \citet{TROTIGNON2006357}. The region within the dashed yellow line shows observations below 250~km identified as the thermosphere region in our analysis. (c) shows the energy spectra of protons within the thermosphere region as per \citet{Halekas2015}.} 
\label{fig:swiaExamples}
\end{figure*}
MAVEN measures in-situ properties of plasma and solar EUV flux it encounters in its flight. We use various in-situ measurements from Solar Wind Ion Analyser (SWIA) \citep{Halekas2015}  and MAGnetometer (MAG) \citep{Connerney2015}, onboard MAVEN  to characterize the influence of solar wind protons and magnetic fields on proton auroras. The details are as follows. 

In its orbit, MAVEN samples plasma from different regions of the magnetosphere --- solar wind upstream of bow shock (SW), magnetosheath (MS) and thermosphere (TH). We identify measurements within SW using the algorithm from \citet{Halekas2017}. We use bow shock and magnetic pile-up boundary positions from \citet{TROTIGNON2006357} to identify measurements within the MS. Finally, we use observations taken below 250 km as TH in-situ observations.  Depending on the altitude, MAVEN does not always sample plasma from SW in each orbit; we only include the orbits in our analysis for which MAVEN samples observations from SW.  Out of the 4130 orbits, after screening for the IUVS limb-scan data availability, SW sampling is available for 2211 MAVEN orbits. This study therefore uses a total of these 2211 orbits for analysis. A distribution of the selected orbits across the considered observation period and is shown in the appendix Figure~\ref{fig:dataDist}. Figure~\ref{fig:swiaExamples}a shows the identified positions of bow shock, magnetic pile-up boundary and TH regions from the SWIA energy spectra for protons for a sample orbit.

The IUVS limb scans are remote sensing measurements and are obtained by accumulating the number of photons along the line-of-sight over a period of approximately 2 minutes. While the in-situ measurement obtained in the regions SW, MS and TH are corresponding to the plasma properties at the location of the spacecraft. Thus, in general, these two measurements correspond to very different regions in the Martian atmosphere. Using the in-situ properties in SW, MS and TH regions allows us to characterize the variability of plasma in these regions to an extent. However, this characterization and therefore our analysis is limited by any local spatial and temporal variations between the locations and measurement times of these in-situ properties and the recorded IUVS limb-scan observations. 

For SW and MS regions, we use the average values of proton and in-situ measurements over an orbit. SW conditions are generally expected to be uniform and therefore orbit-averaged values are a good representative of the solar wind conditions \citep{Ruhunusiri2018,Hughes2021T}. MS plasma environment, spanning over a range of $\rm{\sim 1000~km}$, has a greater variability in plasma-properties and therefore averages may be biased by the presence of outliers. For TH region, we divide all observations within an orbit in 12 equal parts (approximately one for each limb scan) and take the average values from each part. This facilitates characterizing local changes in proton aurora emissions, if present, within an orbit. Each of the 12 parts is marked by corresponding average SZA and altitude. Note that the IUVS limb-scan measurement typically span over the entire $\sim$ 24 minutes of the periapsis pass and therefore the corresponding in-situ measurements do not always confine to the TH region defined above and hence 12 equal-time bins are used instead.

\begin{table*}
\centering
\caption{List of MAVEN/SWIA, MAVEN/MAG and MAVEN/IUVS observations used as input features for modeling Ly-$\alpha$ altitude profiles also observed by MAVEN/IUVS. MAVEN/SWIA and MAVEN/MAG observations are sampled from Upstream solar wind (SW), magnetosheath (MS) and thermosphere (TH) regions, identified as described in Figure~\ref{fig:swiaExamples}. The crustal fields are evaluated at the location of MAVEN using the publicly available model from \citet{Gao2021}.}
\label{tab:features}
\begin{tabular}{|c|c|c|}
\hline
\hline
\multicolumn{2}{|c|}{SWIA In-situ Measurements} & IUVS Remote Sensing Measurements \\
\hline
  {\bf Upstream Solar Wind (SW)}                 &   {\bf Thermosphere (TH:insitu)} & {\bf Thermosphere (TH:co2uvd)}   \\
   IMF Clock Angle ($\mathrm{IMF_{clock}}$)      &  Radial crustal magnetic field ($\mathrm{B_{r,cr}}$)  & co2uvd altitude profile  \\ 
   \cline{3-3}
   IMF Cone Angle ($\mathrm{IMF_{cone}}$)        &  Azimuthal crustal magnetic field ($\mathrm{B_{\phi,cr}}$) & {\bf Thermosphere (TH:rs geom.)}   \\
   MSE-X Proton Speed ($\mathrm{{Vmse}_{x,SW}}$) &  Polar crustal magnetic field ($\mathrm{B_{\theta,cr}}$)  & solar season ($\mathrm{L_s}$)  \\
   Proton temperature ($\mathrm{{T}_{SW}}$)      &  Total magnetic field ($\mathrm{B_{tot,TH}}$)  & local time (lt) \\
   Proton density ($\mathrm{{\rho}_{SW}}$)       & Elevation angle ($\mathrm{\theta_{TH}}$) &  solar zenith angle (SZA) \\
   \cline{1-1}
   {\bf Magnetosheath (MS)}                 & Azimuth angle ($\mathrm{\phi_{TH}}$) & Latitude (lat)   \\
   Total magnetic field ($\mathrm{B_{tot,MS}}$) & MSE-X Proton Speed ($\mathrm{{Vmse}_{x,TH}}$) & Longitude (lon)  \\
   Elevation angle ($\mathrm{\theta_{MS}}$) & MSE-Y Proton Speed ($\mathrm{{Vmse}_{y,TH}}$)  & \\
   Azimuth angle ($\mathrm{\phi_{MS}}$) &  MSE-Z Proton Speed ($\mathrm{{Vmse}_{z,TH}}$) & \\
   MSE-X Proton Speed ($\mathrm{{Vmse}_{x,MS}}$) & Proton temperature ($\mathrm{{T}_{TH}}$) & \\
   MSE-Y Proton Speed ($\mathrm{{Vmse}_{y,MS}}$)  & Proton density ($\mathrm{{\rho}_{TH}}$) & \\
   MSE-Z Proton Speed ($\mathrm{{Vmse}_{z,MS}}$) & Solar Zenith Angle ($\mathrm{{SZA}_{TH}}$)   & \\
   Proton temperature ($\mathrm{{T}_{MS}}$)  &  Altitude ($\mathrm{{ALT}_{TH}}$) & \\
    Proton density ($\mathrm{{\rho}_{MS}}$)  &    & \\
                   \cline{2-2}
                   & {\bf Proton energy spectra (TH:en spec.)} & \\
 \hline
\end{tabular}
\end{table*}
Proton auroras are affected by the solar wind protons, which are monitored by MAVEN/SWIA. SWIA measures proton flux over an energy range of 10~eV -- 5~keV, providing information about proton energies, velocities and temperatures. We convert the velocities provided in Mars-centered Solar Orbital (MSO) coordinate system to Mars Solar Electrical (MSE) coordinate system. In MSE, x points anti-parallel to the solar wind velocity ($\bf{v}_{SW}$), z points in the motional electric field direction $\bf{E}_{SW} = -\bf{v}_{SW} \times \bf{B}_{SW}$ and y completes the right handed system. This conversion from MSO to MSE, although not strictly necessary, is however used in this case to facilitate the inference from our analysis in terms of the plasma flows and currents in the induced magnetosphere of Mars \citep{Ramstad2020}. For MS and TH regions, we use the velocities (converted to MSE), temperature and densities given in the SWIA in-situ key parameters data. In the TH region, velocity, temperature and densities are likely to be overestimated owing to the presence of heavy ions in this region \citep{Halekas2017}(see Section~\ref{sec:shap} for a further explanation of consequences of this bias). Also, we explicitly include the spectrum of protons from the TH region, as an additional input to the model. This spectrum typically shows a peak of flux at the characteristic solar wind energy ${\rm 1~keV}$ when the ENAs converted to penetrating protons are observed \citep{Halekas2015}, An example of the spectrum is shown in Figure~\ref{fig:swiaExamples}c.

In order to study the influence of the magnetic field on proton aurora, we use in-situ measurements of magnetic fields from MAG onboard MAVEN \citep{Connerney2015}. The magnetic-field measurements are provided in MSO coordinate system. For SW measurements, we convert magnetic fields to solar wind clock angle (${\rm B_{clock}}$) and cone angle (${\rm B_{cone}}$) that characterize the direction of Interplanetary Magnetic Field (IMF). For MS and TH regions, we decompose the measurements into the magnitude, elevation angle and azimuth angle \citep{Hara2018}. The elevation angle (${\rm \theta}$) measures how vertical (${\rm \pm~90\degree}$) or horizontal the magnetic field is. The azimuth angle (${\rm \phi}$) measures how east (${\rm 0\degree}$) or north (${\rm 90\degree}$) the horizontal magnetic field is. 

Proton auroras may have preferential occurrence relative to crustal magnetic fields of Mars. These fields are extensively modeled using MAVEN and Mars Global Surveyor (MGS) \citep{Acuna2001} observations. Here, we use a publicly available model from \citet{Gao2021}, that estimates the crustal field with a spatial resolution of ${\rm \sim 200~km}$ and can model MAVEN observations based on these crustal fields within ${\rm \sim nT}$ of the true observations. We use total crustal magnetic field magnitude ${\rm B_{tot,crustal}}$ and spherical co-ordinate system components ${\rm B_{r,crustal}}$, ${\rm B_{\theta,crustal}}$ and ${\rm B_{\phi,crustal}}$ evaluated at the location of the spacecraft using the \citet{Gao2021} model for our analysis. The crustal field spatial resolution of 200~km is comparable to the $\rm {< 300~km}$ scale of the spatial variability observed in patchy proton auroras \citep{Chaffin2022} and therefore may be appropriate for this study.

All observations used as the ANN input are listed in Table~\ref{tab:features}. Since the MS and TH plasma environments are driven by the incident solar wind to a large extent, many measurements, such as magnetic fields, proton speed, density and temperature may be strongly correlated to the corresponding measurements in the SW regions. The input data preparation for developing an efficient ML or ANN model, typically involves selecting features which are independent, and likely to contain the most information about the modeled phenomena. Therefore, it is generally desirable to reduce the inputs to an independent set using dimensional reduction techniques such as Principal Component Analysis \citep{hastie01statisticallearning}. However, one of the key objectives of this study is to explore the possible correlations between various measurements considered in Table~\ref{tab:features} and proton auroras. Hence, we include all features listed in Table~\ref{tab:features} as inputs, without performing any dimensional reduction. Thus, a purpose the Shapley Value analysis presented in Section~\ref{sec:shap} is also feature selection, i.e. identifying and retaining the most important features, for improving the ANN model.

\section{Methods \label{sec:Methods}}
\begin{figure*}[ht!]
\centering
\includegraphics[width=0.75\textwidth]{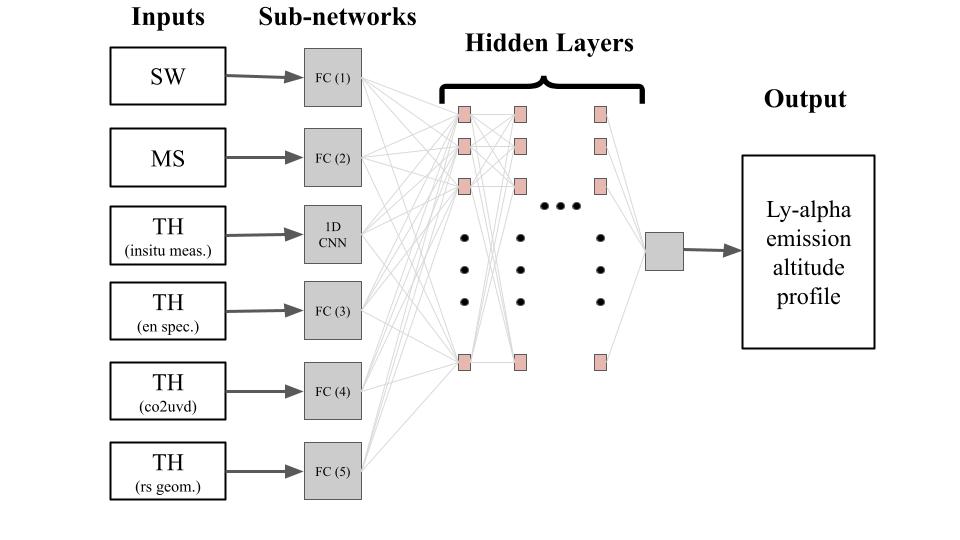}
\caption{The Artificial Neural Network (ANN) Architecture: The ANN takes in SWIA in-situ measurements of proton properties and magnetic fields from upstream solar wind (SW), magnetosheath (MS) and thermosphere (TH) regions, as well as remote sensing geometry measurements of each IUVS limb scan. These input features are summarised in Table~\ref{tab:features}. Fully connected (FC) or 1D Convolutional Neural Network (CNN) sub-networks process the individual set of input features and yield an abstract representation. These representations are further processed by layers of fully connected neurons (hidden layers) to obtain the observed Ly-${\rm \alpha}$ altitude profile of each IUVS limb scan as the output. The details of the FC and 1D-CNN sub-networks, hidden layers and the output layer are given in the main text.}
\label{fig:NN}
\end{figure*}
We build an ANN model for the Ly-${\rm \alpha}$ intensity altitude profiles (between 100 -- 200~km) observed by MAVEN/IUVS in each periapsis limb scan using MAVEN/SWIA in-situ observations of proton energy, density, velocity, temperature, and magnetic fields, modeled crustal magnetic fields, and MAVEN/IUVS observations of co2uvd altitude profiles as well as SZA, ${\rm L_s}$, latitude, longitude, and local time for each limb scan (see Table~\ref{tab:features}). 

\subsection{Artificial neural network architecture}
An ANN is a function ${\rm \bf{Y} = f_{\theta}\left(\bf{X}\right)}$, where 
${\rm \bf{X}}$ are the inputs, ${\rm \bf{Y}}$ is the modeled output and ${\rm \theta}$ are the parameters --- weights and biases --- of the artificial neurons in the network. Each artificial neuron outputs ${\rm \sigma\left(\bf{w}\cdot\bf{x}+\bf{b}\right)}$, where ${\rm \bf{x}}$ is the output from the previous layer (or the input features), ${\rm \bf{w}}$ and ${\rm \bf{b}}$ are the weights and biases of the neuron, and ${\rm \sigma}$ is an activation function \citet{hastie01statisticallearning}. A fully connected (FC) layer of neurons is connected to all neurons (or the inputs) in the previous layer. Layers in a convolution neural network (CNN) are made up of convolution filters, comprising a set of neurons with same weights and biases, that sequentially process outputs from the previous layer (or the inputs) to output feature maps. Convolutional layers are typically used to process images or tabular data. 

In this work, the input features consist different types of observations --- orbit average values (SW and MS), average time series of measurements for each orbit (TH:insitu meas.), energy spectra (TH:en spec.), the co2uvd altitude profiles (TH:co2uvd)  and other remote sensing measurements (TH:rs geom.). Hence, we use different sub-networks to individually process different inputs and obtain a homogeneous abstract representation that is further processed by several FC layers (hidden layers) to yield the output. All inputs except TH:insitu meas. are 1D features and are processed by a FC sub-network with identical architecture. TH:insitu meas. are 2D tabular inputs with 12 average values of each features spaced equally in time during the periapsis scan and are processed by a 1D-CNN. The details of sub-networks FC and 1D-CNN, and FC hidden layers are as follows.

\begin{itemize}
    \item {\bf FC sub-network}: This is made up of three FC layers including the output layer. First and second layers have 128 and 256 neurons respectively while the output layer has 64 neurons. The number of inputs are different for different input groups as listed in Table~\ref{tab:features}. All neurons have the sigmoid activation that gives an output value between 0 and 1 \citet{hastie01statisticallearning}.
    \item {\bf 1D-CNN sub-network}: This comprises of two 1D convolutional layers and one FC layer of neurons. The first convolution layer has 32 filters, while the second convolution layer has 64 filters. The output of the second convolution layer is flattened and fed into a FC layer with 128 neurons. The 1D-CNN operation takes place with a convolution filter of kernel size one and stride one, sliding sequentially over the time series values of each feature, picking up identical patterns from measurements across different times, correlated with the Ly-${\rm \alpha}$ intensity and enhancements. Each neuron in the sub-network has the sigmoid activation. 
    \item {\bf Hidden layers}: The abstract representations of input features are concatenated (vector with length ${\rm 64 \times 5 + 128 = 448}$) and fed into a network of three FC hidden layers with 1024, 512 and 256 neurons respectively. Each neuron in the hidden layers has the sigmoid activation.
    \item {\bf Output layer}: The output of the hidden layers (length=256) is fed into the output layer with 20 neurons modeling the observed Ly-${\rm \alpha}$ emission intensity profile between 100 -- 200 km binned into 20 equally spaced altitude bins. All neuorns in the output layer also have the sigmoid activation function.
\end{itemize}

The ANN architecture is shown in Figure~\ref{fig:NN}. The 1D-CNN sub-network is used to process only the TH:insitu meas. inputs since they are in a 2D tabular form of 17 $\rm{\times}$ 12, corresponding to each equal time bin in the TH region. The 1D-CNN kernel spans one time bin of the TH:insitu meas. and is designed to learn identical patterns across all time bins. All other inputs are 1D and are hence processed with the FC sub-network. The number of hidden layers/convolution filters are typically a power of 2 and also first increase and then decrease. The layer with highest number of neurons contains an abstract encoding of the information learned from inputs. This encoded information is decoded in steps via layers with a decreasing number of neurons to obtain a desired output. The number of neurons in the output layer is dictated by the number of altitude bins for the Ly-$\rm{\alpha}$ profile.

\subsection{Training}
We split the available MAVEN data between October 2014 and April 2022 into three parts for training (60\%), validation (20\%) and test (20\%). The training set data is used to train the ANN i.e. obtain the values of weights and biases that accurately model the output. The validation set data is used to ensure that the model is not overfitting and its performance is generalizable by tuning the hyperparameters (discussed below). A well trained ANN learns concrete patterns from the data to model the output and generalizes to yield good performance on the test set, which is a final test of an ANN model.

Proton aurora observations in different limb scans from an orbit are correlated \citep{Hughes2021T} and, therefore, mixing limb-scan observations from the same orbit in training, validation and test would result in an artificially high performance. Hence, we first randomly split the orbits in the given ratio and all limb scans from each orbit are then included in the respective set. The total number of orbits for training, validation and test data are 1326, 442, and 443 respectively. The total number of limb scans for training, validation and test data are  7178, 2382, and 2384 respectively. 
Note that a random splitting may result in observations from adjacent orbits in the same set. However, from the input features, only ${\rm L_s}$ and latitude evolve slowly and have approximately identical values in adjacent orbits. Other input features, dependent on the solar wind, are expected to vary from orbit to orbit and are known to produce significantly different ${\rm Ly-\alpha}$ response in adjacent orbits (e.g. \citet{Deighan2018}). Hence, presence of a few samples from adjacent orbits in the same set due to a random splitting is not expected to bias the performance significantly.

During training, all samples from the training data are fed into the ANN. For each sample, the ANN returns an output Ly-${\rm \alpha}$ emission intensity altitude profile ${\rm \bf{Y}}$ that is compared with the true observed profile ${\rm \bf{Y_{true}}}$ and a loss function ${\rm L(\bf{Y},\bf{Y}_{true})}$ is computed. The loss function serves as an objective function for optimizing the ANN model parameters \citep{hastie01statisticallearning}. The ANN weights and biases are then modified using a stochastic gradient descent to yield a minimum value for the loss function as the training progresses. For regression problems, such as the one considered here, the most commonly used loss function is a mean squared error (MSE) defined as,
\begin{equation}
    \mathrm{L_{MSE}\left(\bf{Y},\bf{Y}_{true}\right) = \frac{1}{N} \sum_{i}^{N} \frac{1}{M} \sum_{j}^{M} \left(Y_{i}^{j} - Y_{true,i}^{j}\right)^{2},}
\label{eq:mseloss}
\end{equation}
where N is the total number of samples, M=20 is the number of altitude bins and Y's (${\rm Y_{true}}$'s)  are the corresponding Ly-${\rm \alpha}$ intensity values for the modeled (true) scans. The MSE loss function does not explicitly aid the ANN to accurately model the shape of the individual intensity profile, including the characteristic peak for proton auroras. Rather it only aids in reducing the overall mean square error in the intensities. To accurately model the shape, we additionally use structural similarity index (SSIM) defined as,
\begin{widetext}
\begin{equation}
\mathrm{SSIM\left(\bf{Y},\bf{Y}_{true}\right) = \frac{\left(2\mu_{\rm \bf{Y}}\mu_{\rm \bf{Y}_{true}} + C_1\right) \left(2\sigma_{\rm \bf{Y}\bf{Y}_{true}} + C_2\right)}{ \left({\mu^{2}_{\rm \bf{Y}}}+\mu^{2}_{\rm  \bf{Y}_{true}}+C_1\right)+\left(\sigma^{2}_{\rm \bf{Y}}+\sigma^{2}_{\rm \bf{Y}_{true}} + C_2\right)},}
\label{eq:ssim}
\end{equation}
\end{widetext}
where $\rm{\mu}$'s are the means and ${\rm \sigma^{2}}$'s are the variances of ${\rm \bf{Y}}$ and ${\rm \bf{Y}_{true}}$ respectively and $\sigma_{\rm \bf{Y}\bf{Y}_{true}}$ is the covariance. ${\rm C_1}$ and ${\rm C_2}$ are constants with small values used to ensure that the denominator is non-zero. The SSIM is originally designed for comparing images \citep{Wang2004}. Intuitively, the means quantify brightness of images, standard deviations quantify the contrast of images and the covariances are incorporated to compare the local structure of the image \citep{nilsson2020understanding}. The SSIM value ranges between 0 and 1, with the latter meaning that the two profiles are identical. The SSIM loss is therefore defined as,
\begin{equation}
    \mathrm{L_{SSIM}\left(\bf{Y},\bf{Y}_{true}\right) = 1 - SSIM\left(\bf{Y},\bf{Y_{true}}\right).}
\label{eq:ssimloss}
\end{equation}
The EM for each profile provides an additional constraint that can be used for modeling. We use the mean squared error for EM,
\begin{equation}
    \mathrm{L_{MSE,EM}\left(\bf{Y},\bf{Y}_{true}\right) = \frac{1}{N} \sum_{i}^{N} \left(EM\left(Y\right)_{i} - EM\left(Y_{true}\right)_{i}\right)^{2},}
\label{eq:emloss}
\end{equation}
as a third loss function for training. The complete loss function used for training is a weighted sum of the three loss functions
\begin{widetext}
\begin{equation}
    \mathrm{L\left(\bf{Y},\bf{Y}_{true}\right) = L_{MSE}\left(\bf{Y},\bf{Y}_{true}\right) + \lambda_{1} \cdot L_{SSIM}\left(\bf{Y},\bf{Y}_{true}\right) + \lambda_{2} \cdot L_{MSE,EM}\left(\bf{Y},\bf{Y}_{true}\right)},
\label{eq:totloss}
\end{equation}
\end{widetext}
where weight coefficients ${\rm \lambda_1}$ and ${\rm \lambda_2}$ are hyperparameters used for balancing contributions of the different loss function terms.

\begin{figure*}[ht!]
\centering
\includegraphics[width=0.65\textwidth]{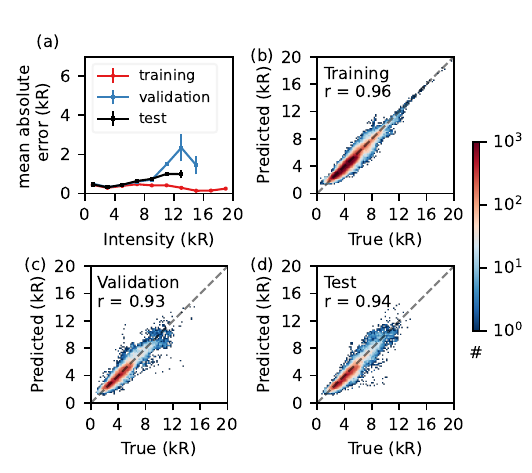}
\caption{Performance of the ANN: (a) Mean absolute error in the predicted intensity as a function of true intensity binned in 10 equal sized bins. The error bars indicate 1${\rm \sigma}$ standard error. (b) (c) \& (d) Heatmaps showing the population of predicted intensity samples binned in 2D as per true and predicted intensities for training, validation, and test sets respectively. The training data is used to obtain the model parameters that minimize the loss function (equation~\ref{eq:totloss}). The validation data is used to ensure that the model performance can generalize to new data and obtain hyperparameters for the training. The test data is totally unseen by the model.  Note that the bins lying closer to the diagonal (dashed line) indicate accurate predictions. The corresponding Pearson correlation values (r) are also noted. The observed intensities of Ly-${\rm alpha}$ emission below $\sim$ 9~kR are accurately reproduced by the ANN model.} 
\label{fig:perTraining}
\end{figure*}

\begin{figure*}[ht!]
\centering
\includegraphics[width=0.6\textwidth]{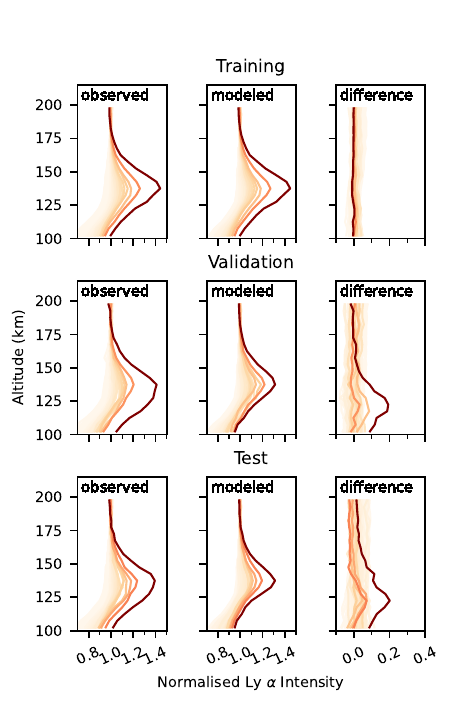}
\caption{Summary of reconstructed Ly-${\rm \alpha}$ intensity altitude profiles for the training, validation and test data. The training data is used to obtain the model parameters that minimize the loss function (equation~\ref{eq:totloss}). The validation data is used to ensure that the model performance can generalize to new data and obtain hyperparameters for the training. The test data is totally unseen by the model. Each profile is a mean profile obtained from a population of profiles binned in percentile bins of the peak intensity enhancements. Darker color indicates increasing enhancement value. Each profile is normalized by mean intensity values between altitude range 160 -- 200~km (after \citet{Hughes2019}). The characteristic shape of the observed proton aurora Ly-${\rm alpha}$ intensity profiles is reproduced reasonably well by the ANN model, except for the cases of extreme intensity enhancements.}
\label{fig:perTrProfiles}
\end{figure*}

Since the ANN learns empirically from the data it is important to have the training data free from any biases with respect to the modeled phenomena i.e. in this case Ly-${\rm \alpha}$ intensity and proton aurora enhancements. As shown in Figure~\ref{fig:iuvsExamples}c, the number of limb scan profiles decrease significantly with increasing enhancement. We bin the Ly-${\rm \alpha}$ profiles in training data as per EM and repeat the number of samples, i.e. oversample, the data in bins with high EM values to match the number in the lowest enhancement bin. Thus, the distribution of Ly-${\rm \alpha}$ profiles to be modeled in the training data is uniform with respect to EM. Oversampling is a standard technique in ML for countering the inadequacy of number of samples in the training data in extreme regimes of the modeled phenomena. However, this may introduce a bias with respect to the inputs to the model. Here, $\rm{L_s}$ and SZA are known to be important for proton auroras and the oversampled training data introduces a sampling bias that the duplicated samples of the higher EMs are dominated by observations from $\rm{L_s \sim 270\degree}$ and ${\rm SZA \sim 50\degree}$ (see appendix Figure~\ref{fig:lsszaosDist}). Consequences of this bias in the oversampled training data are considered in Section~\ref{sec:discussion}.

Each input feature measurement is standardized, by subtracting mean and dividing by standard deviation, before feeding into the ANN. Sine and cosine values of all angular input variables and  local time (with a period of 24~h) are used to ensure continuity across branch cuts.  Each true observed ${\rm Ly-\alpha}$  altitude profile, ${\rm \bf{Y_{true}}}$, is normalized by the maximum ${\rm Ly-\alpha}$ emission value corresponding to each altitude. The sigmoid neurons in the ANN output layer yield a value between 0 and 1, that are reconverted back to the original units using the same observed maximum  ${\rm Ly-\alpha}$ emission values.

Apart from weights and biases, the ANN model also has hyperparameters that are important for training the model and minimizing the loss function. These include the learning rate i.e. the step size for the stochastic gradient descent, batch size i.e. the number of samples used to calculate the gradients for updating weights and biases, the parameters ${\rm \lambda_1}$ and ${\rm \lambda_2}$ for the loss function. We monitor the performance of the model on the validation data to tune the values of these hyperparameters. We implement the ANN model using pytorch \citep{NEURIPS2019_bdbca288}.

\section{Results \label{sec:Results}}
\subsection{Accuracy of the modeled proton auroras \label{sec:performance}}
We use mean absolute error (MAE) of the modeled Ly-$\alpha$ intensities, Pearson correlation between true and modeled intensities as well as EM values obtained from true and modeled profiles for quantifying the accuracy and performance of the ANN model. MAE and a comparison of true and modeled intensities for the training, validation and test data are shown in Figure~\ref{fig:perTraining}. Panel (a) shows MAE in the modeled intensities, calculated for binned populations of the true intensity values. For the training data, MAE is consistently low $\ll$~1~kR across the entire range. In case of the validation and test data, for the true intensities up to 9~kR, MAE remains low $\ll$~1~kR. The MAE increases for the true intensity values $>$ 9~kR to a maximum of $\rm{\sim2~kR}$ and $\rm{\sim1~kR}$ for the validation and test data respectively. Panel (b) and (c) show a heatmap comparison of true and modeled intensity values, with the highest population of samples (indicated in red on the colorbar) lying adjacent to the diagonal. The modeled intensities in the training data are dominated by values up to 9~kR. Higher MAE in the validation and test data for intensities $>$ 9~kR are a consequence of this bias. We find no strong dependence of MAE on $\rm{L_s}$ or SZA as shown in the appendix Figure~\ref{fig:lsszaErr}. The Pearson correlation between the true and modeled intensities yields high values, 0.96, 0.93, and 0.94 for training, validation, and test sets respectively. Similar to the true Ly-$\alpha$ altitude profiles, we obtain the EM values for the modeled Ly-$\alpha$ altitude profiles. Pearson correlations between these EM values, corresponding to the true and modeled Ly-$\alpha$ altitude profiles, are 0.92, 0.65 and 0.60 for training, validation and test sets respectively.

\citet{Hughes2021T} developed a simple linear regression fit for obtaining Ly-$\alpha$ enhancement averaged over an orbit using only the flux of penetrating protons measured by SWIA. They developed three different models for three different scenarios --- nominal conditions, extreme solar events, and events during high dust activity. They obtained ${\rm R^{2}}$ values 0.87, 0.61, 0.43 for these three scenarios respectively. Here, we consider this simple linear regression model by \citet{Hughes2021T} as a baseline for the performance of the ANN model. Compared to \citet{Hughes2021T} simple regression models, the ANN model yields ${\rm R^{2}=0.91}$ for the orbit averaged EMs for the training data. Considering that only the training data is used for the optimization of the ANN model as well as its applicability to all scenarios specified above, the higher ${\rm R^{2}}$ of 0.91 indicates an improved performance over the baseline model. The ANN model also generalizes reasonably well to the validation and test data yielding an ${\rm R^{2}}$ of 0.65 and 0.55 respectively for the orbit averaged EMs.

Figure~\ref{fig:perTrProfiles} shows a direct comparison of mean Ly-$\alpha$ altitude profiles, binned for each percentile population (after \citet{Hughes2019}) as per EM, for the training, validation and test data respectively. The signed difference between the  true and modeled mean profiles for each percentile bin is also shown. Overall, the ANN model accurately reproduces the enhancement shape of the profiles, between altitudes 125 -- 150~km, as expected. For the training data the modeled and observed profiles conform closely to each other across all percentile bins. For the validation and test data, the modeled profiles produce lower peak intensities compared to the true observations for the highest percentile bin. Also, for the highest percentile bins, width of the peak is systematically lower compared to the true profiles. From the difference plots, the modeled intensities, particularly for the highest percentile bins, are lower for altitudes between 110 -- 125~km. For the lower percentile bins, the observed and modeled profiles compare reasonably well.

\subsection{SHAP Values: Identifying important features \label{sec:shap}}
\begin{table}[ht!]
    \centering
    \caption{Ranking of inputs: sum of SHAP values of all measurements in each input group representing an average unsigned (absolute magnitude)} contribution to a Ly-$\alpha$ altitude profile observation.
    \label{tab:shapGlobal}
    \begin{tabular}{cc}
    \hline
    \hline
Input Group & Avg. SHAP Value (kR) \\
\hline
TH:rs meas. & 1.48 \\
TH:co2uvd & 0.79 \\
TH:insitu meas. & 0.33 \\
TH:en spec. & 0.23 \\
MS & 0.14 \\
SW & 0.13 \\
\hline
    \end{tabular}
\end{table}

\begin{table*}
    \centering
    \caption{Selection of important features from upstream solar wind (SW), meagnetosheath (MS) and thermosphere (TH) insitu measurements as well as TH remote sensing geometry measurements. The table lists unsigned (absolute magnitude) SHAP values (normalized for each group by the maximum value) and their Pearson correlation (r) with the corresponding measurements.}
    \label{tab:shapdets1}
    \begin{tabular}{ccc|ccc|ccc}
    \hline
    \hline
        \multicolumn{3}{c|}{\bf TH: res meas.} &  \multicolumn{3}{c|}{\bf TH: insitu meas.} & \multicolumn{3}{c}{\bf MS}\\
    \hline
    {\bf Feature} & {\bf Norm. SHAP} & $\mathrm{\mathbf{r_{shap}}}$ & {\bf Feature} & {\bf Norm. SHAP} & $\mathrm{\mathbf{r_{shap}}}$ & {\bf Feature} & {\bf Norm. SHAP} & $\mathrm{\mathbf{r_{shap}}}$ \\
\hline
$\mathrm{L_s}$      & 0.26 &  0.73                               &  $\mathrm{B_{r,cr}}$       & 0.03 & -0.41  & $\mathrm{B_{tot,MS}}$     & 0.00 &  -0.12 \\
lat           & 0.22 & 0.12                                      &  $\mathrm{B_{\theta,cr}}$  & 0.01 & -0.31  & $\mathrm{\mathbf{\theta_{MS}}}$    & 0.20 &  0.13\\         
lon                 & 0.03 &  -0.67                              &  $\mathrm{B_{\phi,cr}}$    & 0.04 &  0.36  & $\mathrm{\phi_{MS}}$      & 0.16 & 0.07 \\
lt            & 0.15 & -0.73                                     &  $\mathrm{{Vmse}_{x,TH}}$  &  0.09 &  0.76 & $\mathrm{{Vmse}_{x,MS}}$  & 0.00 &  -0.25  \\
SZA     & 0.34 & -0.96                                           &  $\mathrm{{Vmse}_{y,TH}}$  & 0.06 & -0.73                                          & $\mathrm{{Vmse}_{y,MS}}$  & 0.05 & -0.26  \\
\cline{1-3}
\multicolumn{3}{c|}{\bf SW}                                      &  $\mathrm{{Vmse}_{z,TH}}$  & 0.01 & 0.55 & $\mathrm{{Vmse}_{z,MS}}$  & 0.15 & -0.28 \\
\cline{1-3}
{\bf Feature} & {\bf Norm. SHAP} & $\mathrm{\mathbf{r_{shap}}}$  &   $\mathrm{{T}_{TH}}$       & 0.24 & 0.71 & $\mathrm{{T}_{MS}}$ & 0.39 &  -0.35 \\
\cline{1-3}
 $\mathrm{IMF_{clock}}$   & 0.16 &  0.54                         &  $\mathrm{{\rho}_{TH}}$    & 0.00 & -0.21 & $\mathrm{{\rho}_{MS}}$    & 0.05 & -0.44 \\
 $\mathrm{IMF_{cone}}$    & 0.12 & -0.25                         &  $\mathrm{B_{tot,TH}}$     & 0.10 & -0.66 &  \\
 $\mathrm{{Vmse}_{x,SW}}$ & 0.23 &  0.63                         &  $\mathrm{\theta_{TH}}$    & 0.10 & 0.51  &  \\
 $\mathrm{{T}_{SW}}$      & 0.26 &  0.62                         &  $\mathrm{\phi_{TH}}$      & 0.15 &  -0.55 &  \\
 $\mathrm{{\rho}_{SW}}$   & 0.23 &  0.49                         &  $\mathrm{{SZA}_{TH}}$     & 0.15 & 0.59  &                           &      &       \\
                          &      &                               &  $\mathrm{{ALT}_{TH}}$     & 0.02 &  -0.68 &                           &      &       \\
\hline
    
    \end{tabular}

\end{table*}

Complex ML models such as neural networks, although highly efficient and accurate in learning from large high-dimensional datasets, are notoriously difficult to interpret. An interpretation or explanation of a trained ANN model, however, is highly desirable; first and foremost to gain a confidence in applying the model to new data, and subsequently, if possible, to uncover new patterns in the data.  Over the last decade, a number of such interpretation/explanation methods have been developed \citep{Simonyan2013,Zeiler2014,pmlr-v70-sundararajan17a,Selvaraju2017,10.5555/3305890.3306006}. Additive feature attribution methods are a class of explanation models that can be written in the form of a linear function of binary variables that indicate presence/absence of the model inputs. Shapley value, from cooperative game theory, are useful for quantifying impact of each feature on the model output. \citet{NIPS2017_7062} developed SHapley Additive exPlanations (SHAP), based on Shapley values, as a unified measure of feature importance for the class of additive feature attribution methods satisfying a number of desirable properties. SHAP values have proven widely successful and are now a state-of-the-art for explaining a ML model output in terms of contributions of its input features. We use the publicly available python library (\url{https://github.com/shap/shap}) to compute SHAP values (Deep SHAP) for our trained ANN model.

For a given model output, SHAP value for each input is the contribution of the input to the difference between the model output and an expected model output for a set of reference inputs. The sum of SHAP values for all inputs, thus, adds up to the difference between the given model output and expectation value of the reference model output. Here, we use the training data inputs as a reference for calculating the SHAP values. Corresponding to a given set of input measurements, the SHAP values are obtained corresponding to the ${\rm Ly-\alpha}$ emission for each altitude bin in the output between 100 -- 200~km (see Appendix Figure~\ref{fig:shapsSplits}). Since we are interested in understanding the relationship between the inputs and the Ly-$\alpha$ enhancement during proton aurora, we consider mean SHAP values for the Ly-$\alpha$ intensities between altitudes 110 -- 150~km. The ${\rm Ly-\alpha}$ emissions at these altitudes also change due to changes in the neutral H background emission. However, the ANN model is explicitly trained to reconstruct the shape of the Ly-$\rm{\alpha}$ emission at the peak altitudes characteristic of proton auroras via the SSIM (Equation~\ref{eq:ssimloss}) and EM loss (Equation~\ref{eq:ssimloss}) functions. Therefore, the reported SHAPs are expected to capture the contributions of the inputs primarily for proton aurora related enhancements. SHAP values here are measured in kR, same as the intensities. Table~\ref{tab:shapGlobal} lists an average unsigned (absolute magnitude) SHAP value for each group of input features over the training and validation data. The unsigned SHAP values reflect magnitude of the contribution, and therefore importance, of the input feature towards the modeled Ly-${\rm alpha}$ peak altitude intensities. As the table shows, TH:rs geom. and co2uvd intensity profiles that serve as a proxy for the ${\rm CO_2}$ atmosphere (between the altitude range 130 -- 190~km considered here \citet{Deighan2018}) contribute most to the model output on average, while the MS and SW measurements contribute the least. In the following, we analyse SHAP values of important features from each input group in detail.

\subsubsection{In-situ Measurements}
\begin{figure*}[ht!]
\centering
\includegraphics[trim={0.6cm 0.6cm 0.6cm 1cm},clip,width=\textwidth]{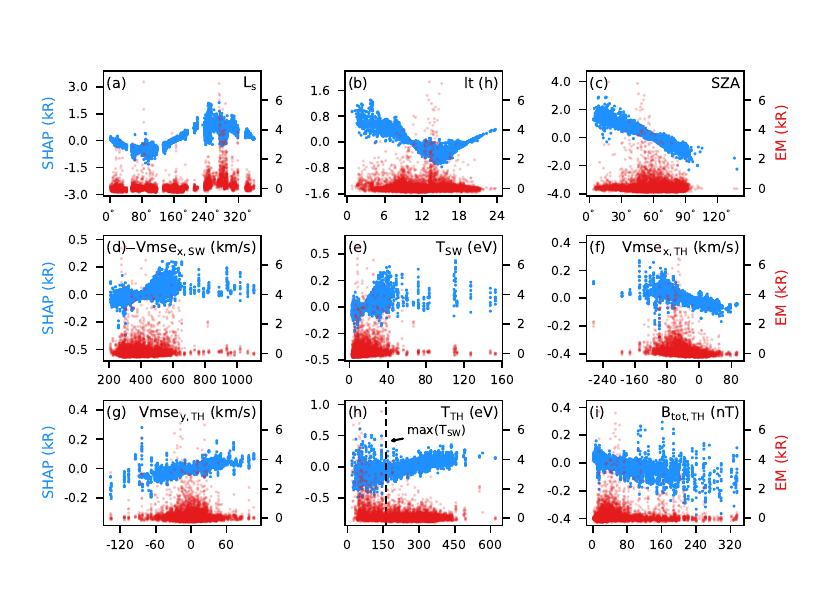}
\caption{Distribution of SHAP values (blue) vs measurements of the important input features from solar wind (SW) and thermosphere (TH) insitu measurements as well as TH remote sensing geometry measurements as identified in Table~\ref{tab:shapdets1}. The distribution of Ly-$\alpha$ enhancement (EM) corresponding to each measurement is also shown (red). The vertical dashed line in Panel (h) indicates the maximum proton temperature in SW region, above which the temperature values in TH region may be strongly influenced by the presence of heavy ions. Note that the vertical scale for SHAPs is different for each panel and Table~\ref{tab:shapdets1} gives a comparison of the relative contributions from each input feature.}  \label{fig:shapGr1}
\end{figure*}

SHAP values for the TH remote sensing and insitu measurements are the highest and third highest respectively. Table~\ref{tab:shapdets1} lists unsigned SHAP values for these features as well as for SW and MS insitu measurements. The SHAP values are normalized by the maximum unsigned value for the respective group. These unsigned SHAP values are an indicator of the contribution of the respective input feature for increasing or decreasing the modeled ${\rm Ly-\alpha}$ intensities. We identify the features from each group as the most significant when the normalized SHAP value exceeds an equal contribution value of ${\rm 1/N}$ where N is the total number of features from the respective group. We also calculate Pearson correlation (r) between the SHAP values and the measured values of the features. High values of r indicate a strong linear relationship between the SHAP value and the feature, suggesting the existence of a definite pattern associated with the Ly-$\alpha$ enhancements. Note that the SHAP values are calculated using $\rm{Ly-\alpha}$ intensities between 110 -- 150~km only where the peak enhancements during proton auroras are typically observed. The positive/negative sign of r indicates the direction of the relationship. The significant normalized SHAP values and/or a relatively high ${\rm r}$ helps us identify input features from these groups, with definite patterns correlated to the modeled Ly-$\alpha$ intensities. Figure~\ref{fig:shapGr1} shows the distribution of SHAP values for these selected features, while additional features also contributing significantly for each group are presented in the appendix Figure~\ref{fig:shapGr2}.

\begin{figure*}[ht]
\centering
\includegraphics[trim={0.25cm 0.25cm 0.25cm 0.25cm},clip,width=\textwidth]{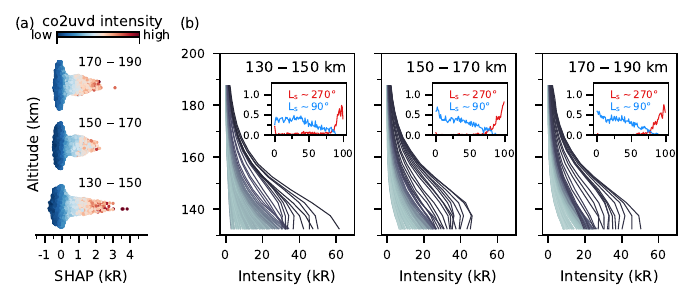}
\caption{SHAP values for the co2uvd altitude profile input. Panel (a) show the SHAP values for co2uvd intensity values grouped in three altitude regions, with the colorbar showing the scale of intensities. Panel (b) shows the mean co2uvd altitude profiles for 100 percentile bins ordered by SHAP values of each altitude range. The darker color corresponds to higher SHAP values. The inset plots show the fraction of observations in each percentile bin corresponding to ${\rm L_s = 270\degree \pm 25\degree}$ and ${\rm L_s=90\degree \pm 25\degree}$. The SHAP values for the co2uvd intensities are highest when the co2uvd intensities are most inflated during the southern summer ${\rm L_s \sim270\degree}$.}
\label{fig:shapCo2uvd}
\end{figure*}
\begin{table}
    \centering
    \caption{Pearson correlation (r) of flux of penetrating protons (observed within the thermosphere, TH) at different energies measured by MAVEN/SWIA, with the flux measured at energy 964.61~eV that is closest to the typical solar wind energy 1~keV. The energies are grouped as per their correlation values.}
    \label{tab:shapdetsEn}
    \begin{tabular}{|cc|cc|}
\hline
    \hline
   \bf{Energy (eV)} & \bf{r} & \bf{Energy (eV)} & \bf{r}\\
\hline
23245.0  & 0.13 & 1720.4   & 0.64\\ 
20114.0  & 0.13 & 1488.7   & 0.64\\ 
\cline{3-4}
17406.0  & 0.15 & 1288.2   & 0.86 \\
15062.0  & 0.15 & 1114.7   & 0.86\\ 
13033.0  & 0.17 & 964.61   & 1.00\\ 
11278.0  & 0.17 & 834.71   & 1.00\\ 
9759.4   & 0.19 & 722.3    & 0.91\\ 
8445.1   & 0.19 & 625.03   & 0.91 \\ 
\cline{3-4}
7307.9   & 0.21 & 540.86   & 0.66\\
6323.7   & 0.21 & 468.02   & 0.66\\ 
\cline{1-2}     
5472.1   & 0.25 & 405.0    & 0.53\\ 
4735.2   & 0.25 & 350.46   & 0.53\\ 
4097.5   & 0.29 & 303.26   & 0.74\\ 
3545.7   & 0.29 & 262.42   & 0.74\\ 
\cline{1-2}     
 3068.2  & 0.42 & 227.08   & 0.77\\
 2655.0  & 0.42 & 196.5    & 0.77\\ 
 2297.5  & 0.50 & 170.04   & 0.73\\ 
 1988.1  & 0.50 & 147.14   & 0.73\\ 
\hline
    \end{tabular}
\end{table}

\begin{table}
    \centering
    \caption{Pearson correlation (r) of flux of penetrating protons (observed within the thermosphere, TH) at different energies measured by MAVEN/SWIA, with the flux measured at energy 964.61~eV that is closest to the typical solar wind energy 1~keV. The energies are grouped as per their correlation values.}
    \label{tab:shapcorrsEn}
    \begin{tabular}{|ccc|}
\hline
    \hline
   \bf{Energy range (eV)} & \bf{Norm. SHAP} & \bf{$\rm{r_{SHAP}}$}\\
\hline
147 -- 541 & 0.27 & 0.06 \\
625 -- 1288 & 0.41 & 0.67 \\
1489 -- 1720 & 0.04 & 0.68 \\
1988 -- 3068 & 0.11 & 0.47 \\
3546 -- 5472 & 0.12 & -0.27 \\
6324 -- 23245 & 0.04 & -0.13 \\
\hline
    \end{tabular}
\end{table}
\begin{figure*}[ht!]
\centering
\includegraphics[trim={0.25cm 0.25cm 0.25cm 0.25cm},clip,width=\textwidth]{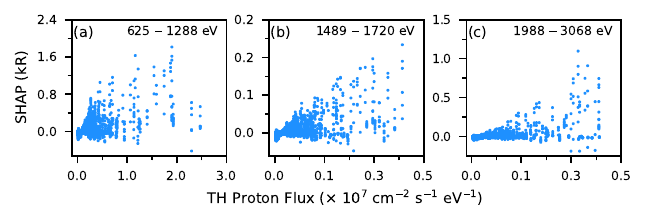}
\caption{Distribution of SHAP values for penetrating proton flux for modeling Ly-$\alpha$. The penetrating proton flux are grouped into different energy bins as per the Pearson correlation of the flux with the flux at $\sim$ 1~keV (see Table~\ref{tab:shapdetsEn}). The contribution of the proton flux near the typical solar wind energy 1~keV is most significant and, overall, the corresponding SHAPs are proportional to the proton flux values in this energy range.}
\label{fig:shapEn}
\end{figure*}

${\rm L_s}$, local time (lt) and SZA for the geometry of remote sensing measurements of Ly-$\alpha$ (and also co2uvd) are the most important features from TH remote sensing measurements. As shown in Figure~\ref{fig:shapGr1}a, from $\rm{L_s = 90\degree}$, i.e. the northern summer solstice, the SHAP values increase approximately linearly overall, and reach a maximum of ${\rm \sim 1.0~kR}$ at approximately $\rm{L_s = 270\degree}$, close to the southern summer solstice. This is consistent with the highest occurrence rates of proton auroras during the southern summer season \citep{Hughes2019}. This is also consistent with the distribution of EM according to $\rm{L_s}$ with the highest enhancements observed during the southern summer season. The SHAP values around $\rm{L_s = 90\degree}$ are negative, and indicating that the ANN model negatively associates those $\rm{L_s}$ values with the ${\rm Ly-\alpha}$ enhancements. The may be a consequence of the bias owing to high occurrences and enhancements in the southern summer season.

From Panel (b), the SHAP values are highest for the early local times (${\rm \sim 2~h}$), with a maximum of ${\rm \sim 0.8~kR}$, and decrease approximately linearly overall until they start increasing again in the evening hours after ${\rm \sim 18~h}$. Panel (c) shows that the SHAP values are highest ${\rm \sim 2.0~kR}$ for low SZAs ${\rm \sim 0\degree}$ and appear to decrease linearly on average with increasing SZAs. The preference of the modeled ${\rm Ly-\alpha}$ enhancements for early hours is noteworthy. However, these early local time values are dominantly sampled from around polar latitudes and periods close to ${\rm L_s = 90\degree}$ and ${\rm L_s = 270\degree}$ during the northern and southern hemisphere summers respectively (see Appendix Figure~\ref{fig:ltBias}) that may be a source of bias. The observed trend in the SHAP values with SZA is consistent with the known proton aurora mechanism, as being a dayside phenomena, the enhancements are expected to be maximum for the lowest SZA \citep{Hughes2019}. The proton aurora enhancements (also shown in the figure) are, however, found to be highest for mid SZAs ${\rm \sim 45\degree}$. This discrepancy is mainly due to the observation bias of the proton aurora occurrence being maximum during southern summer for which the low SZA Ly-$\alpha$ observations are few. Nonetheless, the ANN model here learns the true expected relationship as the Ly-$\alpha$ intensities for different limb scans within a single orbit are also expected to decrease with SZA \citep{Hughes2021T,Fei2023}.

Distribution of SHAP values for the most significant SW measurements ${\rm {Vmse}_{x}}$ and ${\rm T_{SW}}$ are shown in Figure~\ref{fig:shapGr1}, Panels (d) and (e) respectively. The SHAP value variation with both these variables is similar, i.e. generally increasing with increasing solar wind speeds and temperatures. Appendix Figure~\ref{fig:shapGr2}d shows that the SHAP values also increase overall with increasing solar wind densities. Thus, the SHAP contributions for these SW measurements imply increasing modeled Ly-${\rm \alpha}$ peak intensities with increasing solar wind proton flux that is consistent with the known proton aurora mechanism.

Panels (f), (g), (h) and (i) show the SHAP values for TH proton velocities ${\rm Vmse_{x,TH}}$ and ${\rm Vmse_{y,TH}}$, temperature ${\rm T_{TH}}$ and in-situ magnetic field magnitude ${\rm B_{tot,TH}}$. We see an increasing trend in SHAPs with increasing proton speeds in the solar wind direction and ${\rm - mse_{y}}$ direction. The former trend implies increasing ${\rm Ly-\alpha}$ peak intensities with an increasing downward (on the dayside) speed of protons, a relationship that is expected and consistent with the SHAP trends for solar wind proton speed and temperature. The latter relationship may be a result of a strong mutual correlation between the x and y MSE components of the proton speed (e.g. because of the global structure of induced currents \citet{Ramstad2020}). From Panel (h), SHAP values are increasing with increasing values of TH proton temperatures overall, although there appears to be a strong variability at low temperatures ${\rm < 100~eV}$. Proton aurora enhancements are also high in this region of low proton temperatures. The proton temperature values in the TH are likely to be overestimated because of the presence of heavy ions \citep{Halekas2017} and an overall linearly increasing trend of SHAPs for ${\rm T_{TH} > \sim 160~eV}$, the maximum solar wind temperature in our data, may be an artifact. From Panel (i), we find that the SHAP values slowly decrease overall with increasing magnitude of TH in-situ magnetic field and this is also consistent with the distribution of enhancements with respect to these magnetic fields as shown.

\subsubsection{CO2UVD Altitude Profiles}
The co2uvd intensity altitude profiles contribute second highest to the modeled ${\rm Ly-\alpha}$ peak intensities as per the unsigned SHAPs in Table~\ref{tab:shapGlobal}. The co2uvd intensities serve as a proxy for the ${\rm CO_2}$ atmosphere density for altitudes above 130~km. In order to understand the relationship between the modeled Ly-${\rm alpha}$ peak intensities and the co2uvd intensities,  we divide the co2uvd intensity profiles in three groups based on altitudes --- 130 -- 150~km that overlap with the Ly-$\alpha$ peak altitudes, and higher altitudes 150 -- 170~km and 170 -- 190~km. For all three groups, the SHAP values are higher corresponding to the higher intensities at the respective altitudes as shown in Figure~\ref{fig:shapCo2uvd}a. The lowest altitude group yields the highest SHAP values up to $\sim$~4~kR as expected because these include the altitudes for the peak ${\rm Ly-\alpha}$ intensities. The higher altitude groups 150 -- 170~km and 170 -- 190~km yield SHAP values up to $\sim$~2~kR. 

For understanding the relevance of co2uvd intensity values at these different altitude ranges for the modeled Ly-${\rm \alpha}$ emission, we examine co2uvd intensity profiles that are most relevant, i.e. have the highest SHAP values, from each of the three altitude ranges. Figure~\ref{fig:shapCo2uvd}b shows average altitude profiles for 100 percentile bins ordered by the SHAP values for the three altitude groups. The highest SHAP percentile bins for all groups show average co2uvd profiles with higher overall intensities at all altitudes, but particularly at the respective altitude range. The co2uvd intensities as well as the ${\rm CO_2}$ atmosphere are known to vary seasonally with ${\rm L_s}$. The inset plot shows the fraction of samples within each percentile bin approximately corresponding to southern summer solstice (${\rm L_s \sim 270\degree}$) and southern winter solstice (${\rm L_s \sim 90\degree}$). The fraction of profiles from southern summer are higher for the highest SHAP percentile bins 75-100 in all three cases, whereas the fraction of profiles from southern winter systematically decrease to 0 for these SHAP bins.  Thus, SHAP values for the co2uvd intensities at all altitudes are highest when these intensities are most inflated during the southern summer ${\rm L_s \sim270\degree}$. An increase in only the ${\rm CO_2}$ density and therefore co2uvd intensities at these altitudes is, however, expected to increase the absorption of the proton aurora Ly-${\rm \alpha}$ emission \citep{Hughes2021T}. Therefore the relationship of the modeled Ly-${\rm \alpha}$ peak intensities is mainly dictated by the covariant increase in the co2uvd intensities during the southern summer.

\subsubsection{Proton Energy Spectra}
The penetrating proton population is identified by a peak at the characteristic solar wind proton energy ${\rm \sim 1~keV}$ \citep{Halekas2015}. We find that the proton fluxes at other energies are also strongly correlated with the flux at ${\rm \sim 1~keV}$ as shown in Table~\ref{tab:shapdetsEn}. We divide the energy spectra into six energy ranges depending on the degree of correlation with the typical solar wind energy. Table~\ref{tab:shapcorrsEn} lists the normalised SHAP contributions from these energy ranges as well as a Pearson correlation of SHAP with the proton flux values from these ranges. Figure~\ref{fig:shapEn} shows the distribution of SHAP values from the most significant energy ranges with the proton flux at the respective energies. We find that the SHAP values from the energy range 625 -- 1288~eV that contain the typical solar wind proton energy ${\rm \sim 1~keV}$, contribute the most and SHAP values show an increasing trend with the increasing proton flux in this range (Panel (a)) yielding a Pearson correlation of 67\%. SHAP values from higher energies, 1489 -- 1720~eV and 1988 -- 3068~eV are also relatively strongly correlated with the proton flux from these ranges with Pearson correlations of 68\% and 47\% respectively. From Figure~\ref{fig:shapEn}, Panels (b) and (c) respectively, we find that the SHAP values corresponding to these energies increase slowly compared to Panel (a) and are only significant for higher fluxes in these energy ranges. For other energy ranges, shown in Table~\ref{tab:shapcorrsEn}, the SHAP values do not contribute as significantly and also have low Pearson correlation values. The modeled Ly-${\rm \alpha}$ peak altitude intensities therefore are primarily dependent on the proton flux at ${\rm \sim 1~keV}$, the typical energy of the penetrating solar wind protons.

\section{Discussion \label{sec:discussion}}
In this work, we developed an ANN model to obtain the MAVEN/IUVS observed Ly-$\alpha$ altitude profiles that show a marked enhancement at altitudes 110 -- 150~km during proton auroras on Mars. Our ANN model, includes a comprehensive set of in situ measurements from MAVEN/SWIA and MAVEN/MAG to characterise the observed Ly-$\alpha$ emissions in the thermosphere. These measurements included proton densities, temperatures, speeds, magnetic fields, and energies of sampled upstream solar wind, within magnetosheath and also within thermosphere below an altitude of 250~km. These along with a proxy for the ${\rm CO_{2}}$ atmosphere and crustal magnetic fields served as inputs to the ANN model. The trained ANN model reproduces the Ly-$\alpha$ intensities in the validation and test datasets with a high Pearson correlations of 0.93 and 0.94 respectively and mean absolute error ${\rm < 1~kR}$. Although, we did not explicitly model the intensity enhancements (EM), the trained ANN yields a Pearson correlation 0.65 and 0.60 for the validation and test data respectively. The extreme enhancements are not reproduced particularly accurately by the model. These enhancements occur at a lower peak altitude, and typically outside the dominant proton aurora season ${\rm L_s \sim 270\degree}$ , i.e. around the southern summer solstice. The training data is dominated by intense and more frequent proton auroras from the southern summer solstice season, a bias further reinforced by oversampling (see Appendix Figure~\ref{fig:lsszaosDist}).  Hence, the ANN model does not generalize to yield these extreme Ly-$\alpha$ enhancement occurring outside the southern summer solstice season. Mitigation of this bias in the training data e.g. by oversampling extreme enhancements outside the southern summer solstice season may improve the model to yield accurate extreme Ly-$\alpha$ enhancements across all solar seasons.

We performed a SHAP analysis of the trained ANN to explain and validate the modeled Ly-${\rm \alpha}$ intensities and uncover the correlations learned by the ANN between the inputs and the modeled Ly-${\rm \alpha}$ intensity enhancements between altitudes 110 -- 150~km. We find that the SZA (measured at the proton aurora altitudes) and solar season ${\rm L_s}$ contributes most significantly to the modeled Ly-$\alpha$ enhancements. The modeled enhancements peak for low SZAs and ${\rm L_s \sim 270\degree}$ as expected, consistent with the findings of \citet{Hughes2019}. The peak corresponding to ${\rm L_s \sim 270\degree}$ in SHAPs is expected as the training data is dominated by samples from these solar longitudes, which facilitates the ANN to learn the dependence on ${\rm L_s}$. However, relatively few samples from low SZAs are available in the training data (Appendix Figure~\ref{fig:lsszaosDist}) and therefore it is remarkable that the SHAPs trend for SZA shows the true expected relationship with highest contributions from low SZAs $\sim 0\degree$. The observed  ${\rm Ly-\alpha}$ peak enhancements for limb scans within each orbit also increases with decreasing SZA \citep{Hughes2021T}, and the modeled SZA dependence may be a consequence of this pattern in the data. We do not find a significant relationship between the SZA measured at the spacecraft locations and the modeled enhancements (see Appendix Figure~\ref{fig:shapGr2}).

The modeled enhancements are also high for morning local time hours, and peak at very early hours. The observations corresponding to these hours are sampled mainly from around the northern and southern polar region latitudes during the respective summers (Appendix Figure~\ref{fig:ltBias}). Yet, these observations do not seem to be biased by an excessive number of proton aurora occurrences with high EM values (Figure~\ref{fig:shapGr1}). Hence, the significance of these early hours for proton auroras needs to be explored further. 

The modeled enhancements are high for high values of solar wind speed, temperature and also density (see Appendix Figure~\ref{fig:shapGr2}).  Proton auroras are primarily caused by solar wind protons converted to ENAs and therefore the modeled relationship between solar wind speed and density and Ly-${\rm \alpha}$ peak enhancements is expected. \citet{Hughes2023} also found solar wind particle flux and velocity as the primary variables influencing the physics-based models of proton auroras. We, however, do not find significant contributions and relationship between the modeled enhancements and IMF orientation (Appendix Figure~\ref{fig:shapGr2}). Similar to solar wind velocity and temperature, we find an increasing trend in SHAPs for increasing proton temperature, speed measured in the direction of solar wind and speed measured in ${\rm - mse_{y}}$ direction within the thermosphere. The extreme values of these measurements are suspect because of the presence of heavy ions in this region and therefore these trends, particularly for the proton temperature, may be an artifact of this bias.

We find a slow decrease in SHAPs and, therefore the modeled ${\rm Ly-\alpha}$ enhancements between the peak altitudes, with increasing total in-situ magnetic field within the thermosphere. A modeling study by \cite{GERARD2019266} suggested that strong horizontal induced magnetic fields decrease the proton aurora emission intensities. We, however, do not find a strong dependence on the orientation of the thermosphere magnetic field. Yet, from Appendix Figure~\ref{fig:shapGr2}, slightly vertically outward (${\rm \theta_{TH} \sim 30\degree}$) and  westward (${\rm \phi_{TH} \sim 180\degree}$) orientations appear to favour the modeled ${\rm Ly-\alpha}$ enhancements. The in-situ measurements in the magnetosheath region do not yield significant SHAP value contributions or patterns.

The SHAP value analysis solidifies previously known patterns and also uncovers trends that are not directly observed in the data. In the light of the biases identified in the training data, a further analysis and scrutiny is required to verify and establish the relationships presented by the ANN model and SHAP analysis. ${\rm L_s}$ and SZA are the most important contributors, however, they are not primary physical processes, rather only a proxy for conditions during which the interaction of solar wind with Mars is more prominent. Moreover, the magnetosheath and thermosphere in-situ proton measurements are expected to be strongly correlated to the solar wind measurements. Thus, exclusion of ${\rm L_s}$ and SZA, as well as the solar wind measurement inputs, may help the ANN learn subtle correlations of magnetosheath and thermosphere protons, magnetic fields, and crustal fields with the proton aurora occurrences. Training data free of the aforementioned biases is a pre-requirement to obtain robust statistical results from such an exercise. We defer these improvements to future work. An improved ANN model, for the Ly-$\alpha$ intensities or other related phenomena (e.g. ion loss \citet{Fei2023}), can thus be reliably used to simulate, characterise, and model varied Mars--Solar wind interactions for hand-tailored input conditions. Under the paradigm of Physics Informed Neural Networks (PINNs), ANNs can be modeled to explicitly include the known physical processes, further improving their applicability. ANN models, once trained, are considerably inexpensive compared to the traditional numerical models and simulations. Plethora of Mars' magnetosphere, solar wind and atmosphere data available from past and current missions can thus be leveraged using the ever-advancing ML technology, for understanding the dynamics of Mars' magnetosphere and uncovering the mechanisms of the historical loss of Mars' atmosphere.

\section*{Acknowledgements}
This work was supported by the New York University Abu Dhabi
(NYUAD) Institute Research Grants G1502 and CG014, the ASPIRE Award
for Research Excellence (AARE) Grant S1560 by the Advanced
Technology Research Council (ATRC). This work utilized the High
Performance Computing (HPC) resources of NYUAD. We thank the anonymous reviewers for their feedback which helped us in improving the manuscript. We thank Prof. K. R. Sreenivasan for his constant encouragement and support for the project. 

\section*{Author Contributions and Data Availability}
D.B.D and D.A. designed the research and interpreted the results. D.B.D performed the data analysis with contributions from A.H. D.B.D. wrote the paper with contributions from D.A. The MAVEN data used is publicly available at the MAVEN Science Data Center \url{https://lasp.colorado.edu/maven/sdc/public/}.


\begin{thebibliography}{}
\expandafter\ifx\csname natexlab\endcsname\relax\def\natexlab#1{#1}\fi
\providecommand{\url}[1]{\href{#1}{#1}}
\providecommand{\dodoi}[1]{doi:~\href{http://doi.org/#1}{\nolinkurl{#1}}}
\providecommand{\doeprint}[1]{\href{http://ascl.net/#1}{\nolinkurl{http://ascl.net/#1}}}
\providecommand{\doarXiv}[1]{\href{https://arxiv.org/abs/#1}{\nolinkurl{https://arxiv.org/abs/#1}}}

\bibitem[{Acuña {et~al.}(2001)Acuña, Connerney, Wasilewski, Lin, Mitchell,
  Anderson, Carlson, McFadden, Rème, Mazelle, Vignes, Bauer, Cloutier, \&
  Ness}]{Acuna2001}
Acuña, M.~H., Connerney, J. E.~P., Wasilewski, P., {et~al.} 2001, Journal of
  Geophysical Research: Planets, 106, 23403,
  \dodoi{https://doi.org/10.1029/2000JE001404}

\bibitem[{Amiri {et~al.}(2022)Amiri, Brain, Sharaf, Withnell, McGrath,
  Alloghani, Al~Awadhi, Al~Dhafri, Al~Hamadi, Al~Matroushi, Al~Shamsi,
  Al~Shehhi, Chaffin, Deighan, Edwards, Ferrington, Harter, Holsclaw, Kelly,
  Kubitschek, Landin, Lillis, Packard, Parker, Pilinski, Pramman, Reed, Ryan,
  Sanders, Smith, Tomso, Wrigley, Al~Mazmi, Al~Mheiri, Al~Shamsi, Al~Tunaiji,
  Badri, Christensen, England, Fillingim, Forget, Jain, Jakosky, Jones, Lootah,
  Luhmann, Osterloo, Wolff, \& Yousuf}]{Amiri2022}
Amiri, H. E.~S., Brain, D., Sharaf, O., {et~al.} 2022, Space Science Reviews,
  218, 4, \dodoi{10.1007/s11214-021-00868-x}

\bibitem[{Atri {et~al.}(2022)Atri, Dhuri, Simoni, \&
  Sreenivasan}]{atri2022auroras}
Atri, D., Dhuri, D.~B., Simoni, M., \& Sreenivasan, K.~R. 2022, The European
  Physical Journal D, 76, 235

\bibitem[{Bertaux {et~al.}(2005)Bertaux, Leblanc, Witasse, Quemerais,
  Lilensten, Stern, Sandel, \& Korablev}]{Bertaux2005}
Bertaux, J.-L., Leblanc, F., Witasse, O., {et~al.} 2005, Nature, 435, 790,
  \dodoi{10.1038/nature03603}

\bibitem[{Bertaux {et~al.}(2006)Bertaux, Korablev, Perrier, Quémerais,
  Montmessin, Leblanc, Lebonnois, Rannou, Lefèvre, Forget, Fedorova,
  Dimarellis, Reberac, Fonteyn, Chaufray, \& Guibert}]{Bertaux2006}
Bertaux, J.-L., Korablev, O., Perrier, S., {et~al.} 2006, Journal of
  Geophysical Research: Planets, 111,
  \dodoi{https://doi.org/10.1029/2006JE002690}

\bibitem[{Chaffin {et~al.}(2022)Chaffin, Fowler, Deighan, Jain, Holsclaw,
  Hughes, Ramstad, Dong, Brain, AlMazmi, Chirakkil, Correira, England, Evans,
  Fillingim, Lillis, Lootah, Raghuram, McFadden, Halekas, Espley, Schneider,
  Mayyasi, Lee, Curry, \& AlMatroushi}]{Chaffin2022}
Chaffin, M.~S., Fowler, C.~M., Deighan, J., {et~al.} 2022, Geophysical Research
  Letters, 49, e2022GL099881, \dodoi{https://doi.org/10.1029/2022GL099881}

\bibitem[{Connerney {et~al.}(2015)Connerney, Espley, Lawton, Murphy, Odom,
  Oliversen, \& Sheppard}]{Connerney2015}
Connerney, J. E.~P., Espley, J., Lawton, P., {et~al.} 2015, Space Science
  Reviews, 195, 257, \dodoi{10.1007/s11214-015-0169-4}

\bibitem[{Deighan {et~al.}(2018)Deighan, Jain, Chaffin, Fang, Halekas, Clarke,
  Schneider, Stewart, Chaufray, Evans, Stevens, Mayyasi, Stiepen, Crismani,
  McClintock, Holsclaw, Lo, Montmessin, Lef{\`e}vre, \& Jakosky}]{Deighan2018}
Deighan, J., Jain, S.~K., Chaffin, M.~S., {et~al.} 2018, Nature Astronomy, 2,
  802, \dodoi{10.1038/s41550-018-0538-5}

\bibitem[{Gao {et~al.}(2021)Gao, Rong, Klinger, Li, Liu, \& Wei}]{Gao2021}
Gao, J.~W., Rong, Z.~J., Klinger, L., {et~al.} 2021, Earth and Space Science,
  8, e2021EA001860, \dodoi{https://doi.org/10.1029/2021EA001860}

\bibitem[{Goodfellow {et~al.}(2016)Goodfellow, Bengio, \&
  Courville}]{Goodfellow2016}
Goodfellow, I., Bengio, Y., \& Courville, A. 2016, Deep Learning (The MIT
  Press)

\bibitem[{Gérard {et~al.}(2019)Gérard, Hubert, Ritter, Shematovich, \&
  Bisikalo}]{GERARD2019266}
Gérard, J., Hubert, B., Ritter, B., Shematovich, V., \& Bisikalo, D. 2019,
  Icarus, 321, 266, \dodoi{https://doi.org/10.1016/j.icarus.2018.11.013}

\bibitem[{Halekas {et~al.}(2015)Halekas, Taylor, Dalton, Johnson, Curtis,
  McFadden, Mitchell, Lin, \& Jakosky}]{Halekas2015}
Halekas, J.~S., Taylor, E.~R., Dalton, G., {et~al.} 2015, Space Science
  Reviews, 195, 125, \dodoi{10.1007/s11214-013-0029-z}

\bibitem[{Halekas {et~al.}(2017)Halekas, Ruhunusiri, Harada, Collinson,
  Mitchell, Mazelle, McFadden, Connerney, Espley, Eparvier, Luhmann, \&
  Jakosky}]{Halekas2017}
Halekas, J.~S., Ruhunusiri, S., Harada, Y., {et~al.} 2017, Journal of
  Geophysical Research: Space Physics, 122, 547,
  \dodoi{https://doi.org/10.1002/2016JA023167}

\bibitem[{Hara {et~al.}(2018)Hara, Luhmann, Leblanc, Curry, Halekas, Seki,
  Brain, Harada, Mcfadden, DiBraccio, Soobiah, Mitchell, Xu, Mazelle, \&
  Jakosky}]{Hara2018}
Hara, T., Luhmann, J.~G., Leblanc, F., {et~al.} 2018, Journal of Geophysical
  Research: Space Physics, 123, 8572,
  \dodoi{https://doi.org/10.1029/2017JA024798}

\bibitem[{Hastie {et~al.}(2001)Hastie, Tibshirani, \&
  Friedman}]{hastie01statisticallearning}
Hastie, T., Tibshirani, R., \& Friedman, J. 2001, The Elements of Statistical
  Learning, Springer Series in Statistics (New York, NY, USA: Springer New York
  Inc.)

\bibitem[{He {et~al.}(2023)He, Fan, Hughes, Wei, Cui, Schneider, Fraenz, Yao,
  Rong, Chai, Yan, Wu, \& Zhang}]{Fei2023}
He, F., Fan, K., Hughes, A., {et~al.} 2023, Geophysical Research Letters, 50,
  e2023GL102723, \dodoi{https://doi.org/10.1029/2023GL102723}

\bibitem[{Holsclaw {et~al.}(2021)Holsclaw, Deighan, Almatroushi, Chaffin,
  Correira, Evans, Fillingim, Hoskins, Jain, Lillis, Lootah, McPhate, Siegmund,
  Soufli, \& Tyagi}]{Holsclaw2021}
Holsclaw, G.~M., Deighan, J., Almatroushi, H., {et~al.} 2021, Space Science
  Reviews, 217, 79, \dodoi{10.1007/s11214-021-00854-3}

\bibitem[{Hughes {et~al.}(2019)Hughes, Chaffin, Mierkiewicz, Deighan, Jain,
  Schneider, Mayyasi, \& Jakosky}]{Hughes2019}
Hughes, A., Chaffin, M., Mierkiewicz, E., {et~al.} 2019, Journal of Geophysical
  Research: Space Physics, 124, 10533,
  \dodoi{https://doi.org/10.1029/2019JA027140}

\bibitem[{Hughes(2021)}]{Hughes2021T}
Hughes, A.~C. 2021, Doctoral dissertations and master's theses. 611,
  Embry-Riddle Aeronautical University, Daytona Beach, FL.
\newblock \url{https://commons.erau.edu/edt/611}

\bibitem[{Hughes {et~al.}(2023)Hughes, Chaffin, Mierkiewicz, Deighan, Jolitz,
  Kallio, Gronoff, Shematovich, Bisikalo, Halekas, Simon~Wedlund, Schneider,
  Ritter, Girazian, Jain, Gérard, \& Hegyi}]{Hughes2023}
Hughes, A. C.~G., Chaffin, M., Mierkiewicz, E., {et~al.} 2023, Journal of
  Geophysical Research: Space Physics, 128, e2023JA031838,
  \dodoi{https://doi.org/10.1029/2023JA031838}

\bibitem[{Jakosky {et~al.}(2015)Jakosky, Lin, Grebowsky, Luhmann, Mitchell,
  Beutelschies, Priser, Acuna, Andersson, Baird, Baker, Bartlett, Benna,
  Bougher, Brain, Carson, Cauffman, Chamberlin, Chaufray, Cheatom, Clarke,
  Connerney, Cravens, Curtis, Delory, Demcak, DeWolfe, Eparvier, Ergun,
  Eriksson, Espley, Fang, Folta, Fox, Gomez-Rosa, Habenicht, Halekas, Holsclaw,
  Houghton, Howard, Jarosz, Jedrich, Johnson, Kasprzak, Kelley, King, Lankton,
  Larson, Leblanc, Lefevre, Lillis, Mahaffy, Mazelle, McClintock, McFadden,
  Mitchell, Montmessin, Morrissey, Peterson, Possel, Sauvaud, Schneider,
  Sidney, Sparacino, Stewart, Tolson, Toublanc, Waters, Woods, Yelle, \&
  Zurek}]{Jakosky2015}
Jakosky, B.~M., Lin, R.~P., Grebowsky, J.~M., {et~al.} 2015, Space Science
  Reviews, 195, 3, \dodoi{10.1007/s11214-015-0139-x}

\bibitem[{Lillis {et~al.}(2022)Lillis, Deighan, Brain, Fillingim, Jain,
  Chaffin, England, Holsclaw, Chirakkil, Al~Matroushi, Lootah, Al~Mazmi,
  Thiemann, Eparvier, Schneider, \& Curry}]{Lillis2022}
Lillis, R.~J., Deighan, J., Brain, D., {et~al.} 2022, Geophysical Research
  Letters, 49, e2022GL099820, \dodoi{https://doi.org/10.1029/2022GL099820}

\bibitem[{Lundberg \& Lee(2017)}]{NIPS2017_7062}
Lundberg, S.~M., \& Lee, S.-I. 2017, in Advances in Neural Information
  Processing Systems 30, ed. I.~Guyon, U.~V. Luxburg, S.~Bengio, H.~Wallach,
  R.~Fergus, S.~Vishwanathan, \& R.~Garnett (Curran Associates, Inc.),
  4765--4774.
\newblock
  \url{http://papers.nips.cc/paper/7062-a-unified-approach-to-interpreting-model-predictions.pdf}

\bibitem[{McClintock {et~al.}(2015)McClintock, Schneider, Holsclaw, Clarke,
  Hoskins, Stewart, Montmessin, Yelle, \& Deighan}]{McClintock2015}
McClintock, W.~E., Schneider, N.~M., Holsclaw, G.~M., {et~al.} 2015, Space
  Science Reviews, 195, 75, \dodoi{10.1007/s11214-014-0098-7}

\bibitem[{Nilsson \& Akenine-Möller(2020)}]{nilsson2020understanding}
Nilsson, J., \& Akenine-Möller, T. 2020.
\newblock \doarXiv{2006.13846}

\bibitem[{Paszke {et~al.}(2019)Paszke, Gross, Massa, Lerer, Bradbury, Chanan,
  Killeen, Lin, Gimelshein, Antiga, Desmaison, Kopf, Yang, DeVito, Raison,
  Tejani, Chilamkurthy, Steiner, Fang, Bai, \& Chintala}]{NEURIPS2019_bdbca288}
Paszke, A., Gross, S., Massa, F., {et~al.} 2019, in Advances in Neural
  Information Processing Systems, ed. H.~Wallach, H.~Larochelle,
  A.~Beygelzimer, F.~d\textquotesingle Alch\'{e}-Buc, E.~Fox, \& R.~Garnett,
  Vol.~32 (Curran Associates, Inc.).
\newblock
  \url{https://proceedings.neurips.cc/paper_files/paper/2019/file/bdbca288fee7f92f2bfa9f7012727740-Paper.pdf}

\bibitem[{Ramstad {et~al.}(2020)Ramstad, Brain, Dong, Espley, Halekas, \&
  Jakosky}]{Ramstad2020}
Ramstad, R., Brain, D.~A., Dong, Y., {et~al.} 2020, Nature Astronomy, 4, 979,
  \dodoi{10.1038/s41550-020-1099-y}

\bibitem[{Ritter {et~al.}(2018)Ritter, Gérard, Hubert, Rodriguez, \&
  Montmessin}]{Ritter2018}
Ritter, B., Gérard, J.-C., Hubert, B., Rodriguez, L., \& Montmessin, F. 2018,
  Geophysical Research Letters, 45, 612,
  \dodoi{https://doi.org/10.1002/2017GL076235}

\bibitem[{Ruhunusiri {et~al.}(2018)Ruhunusiri, Halekas, Espley, Eparvier,
  Brain, Mazelle, Harada, DiBraccio, Dong, Ma, Thiemann, Mitchell, \&
  Jakosky}]{Ruhunusiri2018}
Ruhunusiri, S., Halekas, J.~S., Espley, J.~R., {et~al.} 2018, Geophysical
  Research Letters, 45, 10,855, \dodoi{https://doi.org/10.1029/2018GL079282}

\bibitem[{Schneider {et~al.}(2015)Schneider, Deighan, Jain, Stiepen, Stewart,
  Larson, Mitchell, Mazelle, Lee, Lillis, Evans, Brain, Stevens, McClintock,
  Chaffin, Crismani, Holsclaw, Lefevre, Lo, Clarke, Montmessin, \&
  Jakosky}]{Schneider2015}
Schneider, N.~M., Deighan, J.~I., Jain, S.~K., {et~al.} 2015, Science, 350,
  aad0313, \dodoi{10.1126/science.aad0313}

\bibitem[{Selvaraju {et~al.}(2017)Selvaraju, Cogswell, Das, Vedantam, Parikh,
  \& Batra}]{Selvaraju2017}
Selvaraju, R.~R., Cogswell, M., Das, A., {et~al.} 2017, 2017 IEEE International
  Conference on Computer Vision (ICCV), 1, 618

\bibitem[{Shrikumar {et~al.}(2017)Shrikumar, Greenside, \&
  Kundaje}]{10.5555/3305890.3306006}
Shrikumar, A., Greenside, P., \& Kundaje, A. 2017, in Proceedings of the 34th
  International Conference on Machine Learning - Volume 70, ICML'17 (JMLR.org),
  3145–3153

\bibitem[{Simonyan {et~al.}(2013)Simonyan, Vedaldi, \&
  Zisserman}]{Simonyan2013}
Simonyan, K., Vedaldi, A., \& Zisserman, A. 2013, CoRR, abs/1312.6034, 1

\bibitem[{Sundararajan {et~al.}(2017)Sundararajan, Taly, \&
  Yan}]{pmlr-v70-sundararajan17a}
Sundararajan, M., Taly, A., \& Yan, Q. 2017, in Proceedings of Machine Learning
  Research, Vol.~70, Proceedings of the 34th International Conference on
  Machine Learning, ed. D.~Precup \& Y.~W. Teh (PMLR), 3319--3328.
\newblock \url{https://proceedings.mlr.press/v70/sundararajan17a.html}

\bibitem[{Trotignon {et~al.}(2006)Trotignon, Mazelle, Bertucci, \&
  Acuña}]{TROTIGNON2006357}
Trotignon, J., Mazelle, C., Bertucci, C., \& Acuña, M. 2006, Planetary and
  Space Science, 54, 357, \dodoi{https://doi.org/10.1016/j.pss.2006.01.003}

\bibitem[{Wang {et~al.}(2004)Wang, Bovik, Sheikh, \& Simoncelli}]{Wang2004}
Wang, Z., Bovik, A., Sheikh, H., \& Simoncelli, E. 2004, IEEE Transactions on
  Image Processing, 13, 600, \dodoi{10.1109/TIP.2003.819861}

\bibitem[{Zeiler \& Fergus(2014)}]{Zeiler2014}
Zeiler, M., \& Fergus, R. 2014, in Lecture Notes in Computer Science (including
  subseries Lecture Notes in Artificial Intelligence and Lecture Notes in
  Bioinformatics), Vol. 8689 LNCS, Computer Vision, ECCV 2014 - 13th European
  Conference, Proceedings, part 1 edn. (Springer Verlag), 818--833,
  \dodoi{10.1007/978-3-319-10590-1_53}

\end{thebibliography}

\appendix
\section{Data Distribtion in training, validation and test sets}

\begin{figure*}[ht!]
\centering
\includegraphics[trim={0.3cm 0.3cm 0.3cm 0.2cm},clip,width=\textwidth]{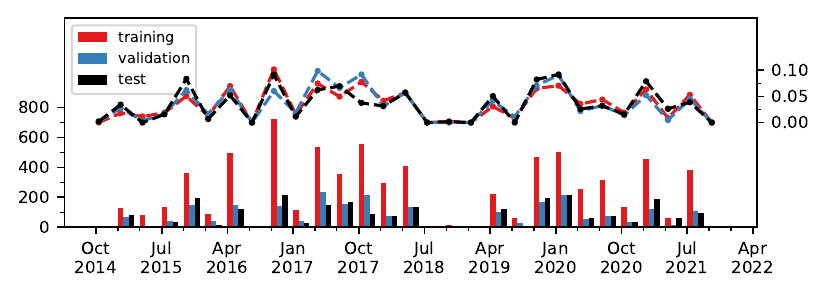}
\caption{Number of samples in training, validation and test dataset across the considered time period for MAVEN observations between Oct 2014 --- April 2022. The size of each bin is three months. The dashed lines show fractional data count in each bin.}  \label{fig:dataDist}
\end{figure*}

\begin{figure*}[ht!]
\centering
\includegraphics[trim={0.3cm 0.3cm 0.3cm 0.2cm},clip,width=\textwidth]{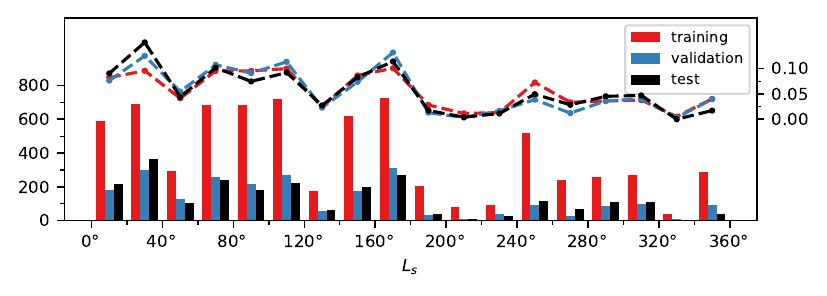}
\caption{Number of samples in training, validation and test dataset across the solar longitude $L_s$ during the considered observation period. The size of each bin each 20$\degree$. The dashed lines show fractional data count in each bin.}  \label{fig:lsDist}
\end{figure*}

\begin{figure*}[ht!]
\centering
\includegraphics[trim={0.3cm 0.3cm 0.3cm 0.2cm},clip,width=\textwidth]{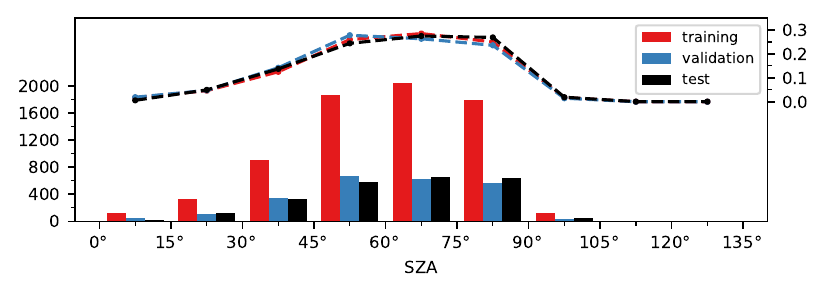}
\caption{Number of samples in training, validation and test dataset across the considered solar zenith angle SZA. The size of each bin is 15$\degree$. The dashed lines show fractional data count in each bin.}  \label{fig:szaDist}
\end{figure*}

\begin{figure*}[ht!]
\centering
\includegraphics[trim={0.3cm 0.3cm 0.3cm 0.2cm},clip,width=\textwidth]{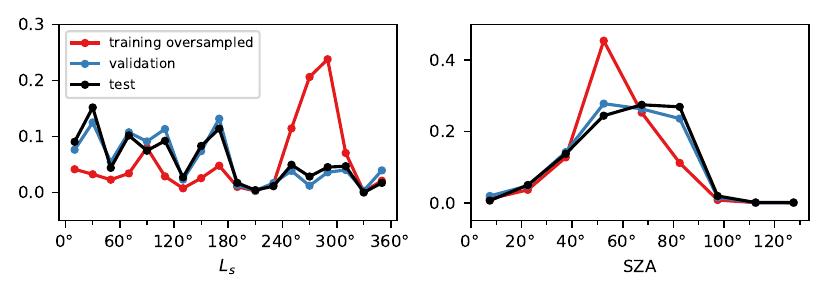}
\caption{Comparison of the fraction of number of samples in training dataset after oversampling with validation and test dataset for the considered solar longitude ${\rm L_s}$ and solar zenith angle SZA.}  \label{fig:lsszaosDist}
\end{figure*}

\begin{figure*}[ht!]
\centering
\includegraphics[trim={0.3cm 0.3cm 0.3cm 0.2cm},clip,width=\textwidth]{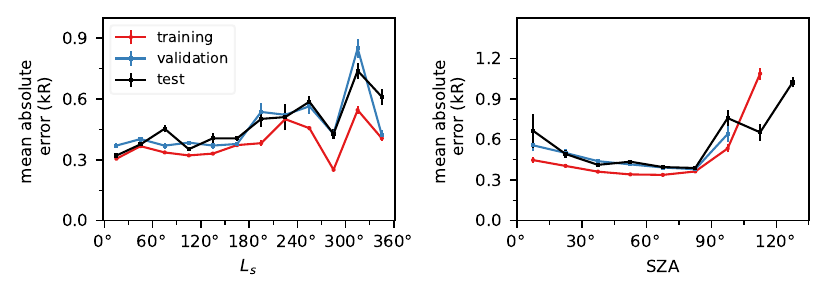}
\caption{Comparison mean squared error in the modeled intensities in training dataset (after oversampling) with validation and test dataset for the considered solar longitude $L_s$ and solar zenith angle SZA.}  \label{fig:lsszaErr}
\end{figure*}

\clearpage
\section{SHAP analysis}
\begin{figure*}[ht!]
\centering
\includegraphics[trim={1.5cm 1.8cm 1.5cm 2.2cm},clip,width=0.7\textwidth]{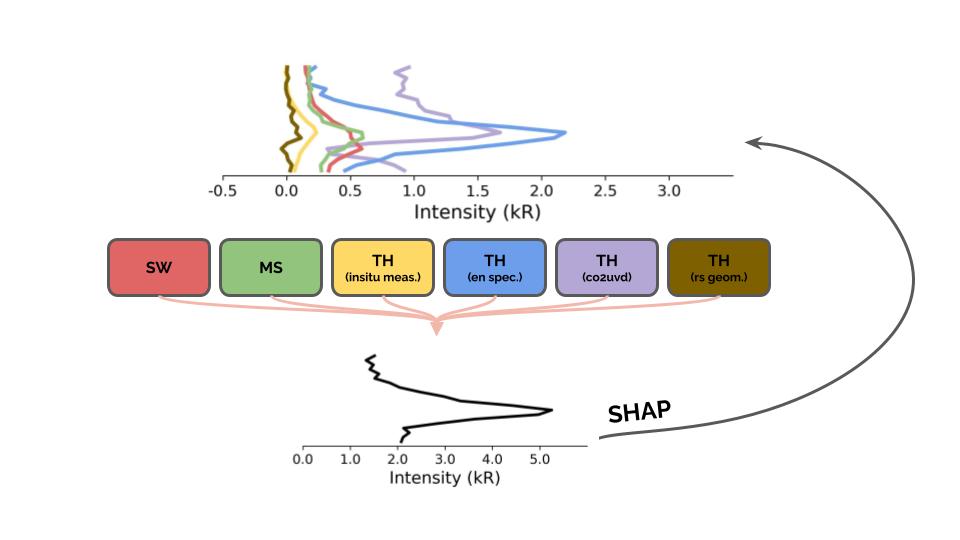}
\caption{An illustration of SHAPs: The SHAP (DeepSHAP) values altitude profiles splitting the ANN outputs into contributions from the input feature groups. The altitude profiles are color coded as per the legend.}  \label{fig:shapsSplits}
\end{figure*}

\begin{figure*}[ht!]
\centering
\includegraphics[trim={0.6cm 0.6cm 0.6cm 1cm},clip,width=\textwidth]{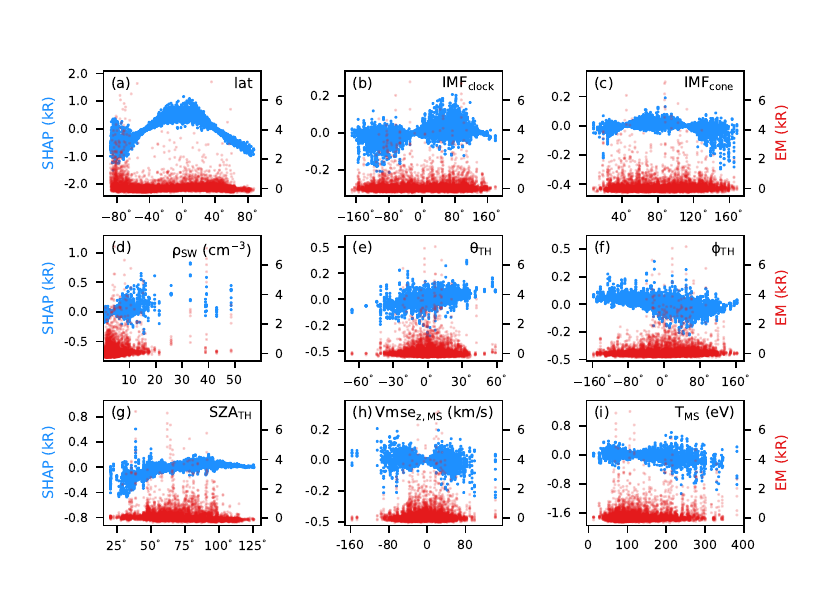}
\caption{Distribution of SHAP values (blue) vs measurements of the important input features from meagnetosheath (MS) and thermosphere (TH) insitu measurements as well as TH remote sensing geometry measurements as identified in Table~\ref{tab:shapdets1}. The distribution of Ly-$\alpha$ enhancement (EM) corresponding to each measurement is also shown (red).}  \label{fig:shapGr2}
\end{figure*}

\begin{figure*}[ht!]
\centering
\includegraphics[trim={0.3cm 0.3cm 0.3cm 0.2cm},clip,width=\textwidth]{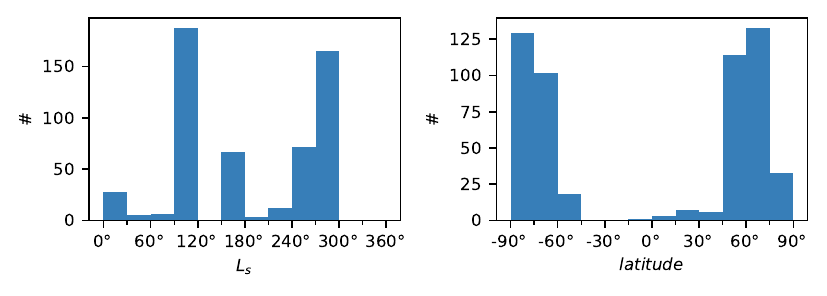}
\caption{Distribution of counts for $\rm{Ly-\alpha}$ observations with local time $<$ 6~h with respect to solar longitude ${\rm L_s}$ and latitude.}  \label{fig:ltBias}
\end{figure*}
\begin{figure*}[ht!]
\centering
\includegraphics[trim={0.3cm 0.3cm 0.3cm 0.2cm},clip,width=\textwidth]{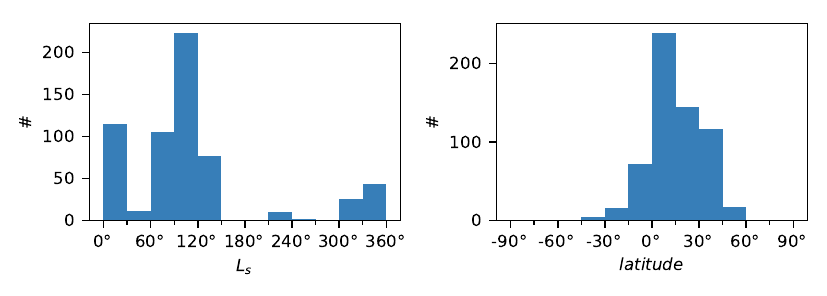}
\caption{Distribution of counts for $\rm{Ly-\alpha}$ observations with SZA $<$ $30\degree$ with respect to solar longitude ${\rm L_s}$ and latitude.}  \label{fig:szaBias}
\end{figure*}

\end{document}